\documentclass[twocolumn,showpacs,showkeys,preprintnumbers,nofootinbib,amsmath,amssymb,eqsecnum]{revtex4}

\def\ba{\begin{eqnarray}}
\def\ea{\end{eqnarray}}
\def\Ep#1{Eq.\ (\ref{#1})}

\def\EQN#1{\label{#1}}
\def\ket{\rangle}
\def\bra{\langle}

\def\>{\rangle}
\def\<{\langle}
\def\no{\nonumber\\}
\def\dg{\dagger}
\def\lam{\lambda}
\def\al{\alpha}
\def\bt{\beta}

\def\ome{\omega}
\def\del{\delta}

\def\gam{\gamma}

\newcommand{\w}{\omega}
\newcommand{\wb}{\bar{\omega}}
\newcommand{\wt}{\tilde{\omega}}

\newcommand{\Ab}{\bar{A}}

\newcommand{\Db}{\bar{D}}

\newcommand{\eps}{\epsilon}
\newcommand{\At}{\tilde{A}}
\newcommand{\pd}{\partial}
\newcommand{\cb}{\bar{c}}
\newcommand{\db}{\bar{d}}
\newcommand{\Bb}{\bar{B}}
\newcommand{\ct}{\tilde{c}}
\newcommand{\dt}{\tilde{d}}
\newcommand{\Bt}{\tilde{B}}
\newcommand{\Dt}{\tilde{D}}
\newcommand{\vt}{\tilde{v}}
\newcommand{\Yt}{Y}
\newcommand{\Xt}{X}
\newcommand{\Pito}{\Pi}
\newcommand{\Pih}{\hat{\Pi}}
\newcommand{\Po}{P}
\newcommand{\Yb}{\bar{Y}}
\newcommand{\Xb}{\bar{X}}
\newcommand{\phr}{|\phi\rangle}
\newcommand{\phl}{\langle\phi|}
\newcommand{\rph}{\phi\rangle}
\newcommand{\lph}{\langle\phi}

\newcommand{\qpr}{|{x'}_1\rangle}
\newcommand{\ql}{\langle x_1|}

\newcommand{\pdi}{\partial}
\newcommand{\Pibo}{\bar{\Pi}}
\newcommand{\thetao}{\theta}
\newcommand{\rhoo}{{\tilde\rho}}

\begin{document}

\title{ Exact Markovian kinetic equation for a quantum Brownian oscillator}
\author{ B. A. Tay}
\altaffiliation{Current address: 143, Jalan Green, 93150 Kuching,
Sarawak, Malaysia.}

\author{G. Ordonez }
\altaffiliation{Current address: Butler University, 4600 Sunset Ave., Indianapolis, IN 46208.}
\email{gonzalo@ph.utexas.edu} \affiliation{Center for Studies in
Statistical Mechanics and
Complex Systems,\\
The University of Texas at Austin,\\
        1 University Station C1609 Austin, TX 78712 U.S.A.
        }


\date{Phys. Rev. E {\bf{73}} 016120 (2006)}

\begin{abstract}

We derive an exact Markovian kinetic equation for an oscillator linearly coupled to a heat bath,
describing quantum Brownian motion. Our work is based on the
subdynamics formulation developed by Prigogine and collaborators.
The space of distribution functions is decomposed into independent
subspaces that  remain invariant under Liouville dynamics. For
integrable systems in Poincar\'e's sense the invariant subspaces
follow the dynamics of uncoupled, renormalized particles. In
contrast for non-integrable systems,  the invariant subspaces follow
a dynamics with broken-time symmetry, involving generalized
functions.  This result indicates that irreversibility and
stochasticity are exact properties of dynamics in generalized
function spaces. We  comment on the relation between our Markovian
kinetic equation and the Hu-Paz-Zhang equation.

\pacs{02.50.Ga, 05.40.Jc, 05.30.Ch}

\end{abstract}

\keywords{Brownian motion, irreversibility, Markovian equations,
integrability}

\maketitle

 \section{Introduction}

A well-known model of quantum Brownian motion is a harmonic oscillator linearly coupled to
a bath of field modes.  The
Hamiltonian is  (with $\hbar =1$)
\ba  \label{HHPZ}
        H  &=& \frac{1}{2 M_1}  {{p}}^2_1 +\frac{m_1}{2}\w_1^2 {q}_1^2  +
        \sum_{k=0}^\infty  \frac{1}{2 M_k}  {{p}}^2_k +\frac{m_k}{2}\w_k^2 {q}_k^2 \nonumber\\
        &+&  \lam \sum_{k=0}^\infty C_k {q}_1 {q}_k,
\ea
where ${q}_1, {p}_1$ are  the positions and momenta of the harmonic
oscillator and ${q}_k, {p}_k$ are positions and momenta of the field
oscillators. Here $k$ are the wave numbers, and $\lam$ is a dimensionless coupling constant.\footnote{Note that for a finite system the wave numbers $k$ are not integers (see \Ep{knL}). Thus the  bath variables  (e.g. $q_k$ and $p_k$) , do not take the ``$1$'' index reserved for the harmonic oscillator variables $q_1$, $p_1$.}

The Hamiltonian (\ref{HHPZ}) has been considered in numerous papers (see Dekker's review \cite{Dekker}). Hu, Paz and Zhang have obtained an exact equation for the reduced
density matrix of the oscillator using a path-integral method,
\ba    \label{HPZ92}
         i\frac{\pd}{\pd t}  \rho_r &=& \left[
                            -\frac{\w_1}{2}\left(\frac{\pdi^2}{\pdi x^2_1}
              -\frac{\pdi^2}{\pdi{x'}^2_1}\right)  + \frac{\tilde{\Omega}^2 (t) }{2\w_1} (x^2_1-{x'}^2_1)
              \right] \rho_r \no
             &-& i \, \Gamma(t) (x_1-x'_1)\left(\frac{\pdi}{\pdi x_1}
              - \frac{\pdi}{\pdi{x'}_1}\right) \rho_r \no
              &-& \frac{i}{\w_1} \Gamma(t) h(t)
                (x_1-x'_1)^2  \rho_r  \no
                &+& \Gamma(t) f(t) (x_1 - x'_1 ) \left(\frac{\pdi}{\pdi x_1}
             + \frac{\pdi}{\pdi{x'}_1}\right) \rho_r,
\ea
where
\ba \label{x}
        x_1=\sqrt{M_1 \w_1} \, q_1,  \qquad
{x'}_1=\sqrt{M_1 \w_1} \, {q'}_1 \ea
are dimensionless coordinates, and  the time-dependent coefficients
are defined in Ref. \cite{HPZ}.

The  terms with $\Gamma(t)$ and $\Gamma(t) h(t)$ on the right hand
side of \Ep{HPZ92} suggest the existence of damping and diffusion
processes characteristic of Brownian motion. Strictly speaking
though, the Hu-Paz-Zhang (HPZ) equation (\ref{HPZ92}) is
time-reversal invariant and it corresponds to a deterministic
evolution of wave functions. Indeed,   the solution of the HPZ
equation is  equivalent to the solution obtained from
Schr\"odinger's equation, i.e.,
\ba \label{HS}
       \rho_r(t) = {\rm Tr_F} \left( e^{-iHt} \rho(0) e^{iHt} \right)
\ea
where $ {\rm Tr_F}$ means trace over the field and
\ba \label{rhoz}
      \rho(0) =  \sum_{\alpha}  \rho_\alpha
      |\psi_\alpha \ket\bra \psi_\alpha|
\ea
is the initial density matrix, diagonalized in a suitable basis of
wave functions $|\psi_\alpha\ket$. The wave functions
$|\psi_\alpha\ket$ form a complete orthonormal basis of the whole
system of harmonic oscillator and heat bath.
 Due to the equivalence (\ref{HS}),  the HPZ equation  describes a time-reversible, deterministic evolution of each wave function. This contrasts with  true Brownian motion, described by a Markovian equation with broken time-symmetry, which corresponds to a stochastic evolution of each wave function $|\psi_\alpha\ket$ \cite{Weiss}.

The derivation of irreversible Markovian equations from dynamics has
been a great challenge \cite{Balescu}. This is related to the
apparent  incompatibility between the second law of thermodynamics
and time-reversible dynamics. One point of view is that Markovian
equations appear as an approximation of the dynamical equations.
This is the so-called Markovian approximation, valid for weak
coupling  between interacting particles, and for time scales of the
order of the relaxation time to equilibrium \cite{spohn}.

However, one can take a  different point of view, where Markovian
equations are formally derived from dynamical equations without any
approximation. This is the {\it subdynamics} formulation  developed
by Prigogine and collaborators  \cite{Balescu,Subd, G73, G85,
Petrosky89, Petrosky97,TB}. In this approach, essential elements
are the distinction between integrable and non-integrable systems in
the sense of Poincar\'e, and the use of generalized functions
\cite{PPT}. In this paper we  derive an exact Markovian equation for
the quantum Brownian oscillator, based on this approach.  This equation
is valid  for both weak and strong coupling. As we will show, for
weak coupling, it agrees with the HPZ equation.

A few exact results using subdynamics have already been obtained
\cite{Balescu, deHaan}. However, to our knowledge, there was no
derivation of an exact Markovian equation for quantum Brownian
motion. Previous formulations were centered on density operators.
Here we focus on the observables, that is, on products of creation
and annihilation operators. This allows us to consider arbitrary
N-particle sectors in a non-perturbative way.

This paper is organized as follows. In Section \ref{Pisub} we
introduce our formulation of subdynamics. As in the original
formulation, we introduce the projection super operators $\Pito$ and
$\Pih=1-\Pito$, which define invariant subspaces of the Liouville
super operator. In Sec. \ref{IntNoInt} we define integrability and
non-integrability in Poincar\'e's sense. In Secs.
\ref{PiInt}-\ref{Recurrence}, we construct the $\Pi$ projector for
the integrable case and derive recursive relations for this
projector. Extending these relations we construct $\Pito$ for the
the non-integrable case in Sec. \ref{PiNoInt}. This leads to our
Markovian equation in Sec. \ref{Exact}. In Sec. \ref{HPZKE} we
compare this equation with the HPZ equation. Concluding remarks are
presented in Sec. \ref{Conclusions}.  Additional calculations are
presented in the Appendices.

\section{Subdynamics}
\label{Pisub}

In this section we introduce the main ideas of our approach.  We
focus on the quantum Brownian oscillator model.

We  consider a one dimensional space. We start with  the system in a
box of size $L$ and impose periodic boundary conditions. Then in
\Ep{HHPZ} we have
\ba\label{knL}
k = 2 \pi n/L
\ea
 with integer $n$.  We are
interested in the limit $L\to\infty$, where the spectrum of field
frequencies $\w_k$ becomes continuous. We will assume that $C_k =
C_{-k}$ and $\w_k = \w_{-k}$. This allows us to to restrict $k\ge0$
keeping only the symmetric part of the ${q}_k$ operators, i.e., we
set ${q}_k = {q}_{-k}$. We will assume as well that there is no
degeneracy in the spectrum of $\w_k$ for $k\ge0$.

It will be convenient to express the Hamiltonian (\ref{HHPZ})  in
terms of annihilation and creation operators. We express the
coordinates $q_i$  as
\ba  \label{qHPZ}
        {q}_i = \frac{1}{\sqrt{2 M_i \w_i}} (a_i + a_i^\dg)
        \,,   \qquad i= 1, \, k
        \,,
\ea where
$a^\dagger_i$, $a_i$ are bosonic creation and annihilation operators
of the particle ($i=1$) and field ($\{i\}=\{k\}$). These operators
satisfy the usual commutation relations
\ba
        [a_i,\,a_j^\dg]=\delta_{ij}  \,,
\ea
for $i,j = 1$ or $k$.  For the momenta we have
\ba  \label{pHPZ}
        {p}_i = -i\sqrt{\frac{M_i \w_i}{2  }} (a_i - a_i^\dg)
        \,,   \qquad i= 1, \, k
        \,.
\ea
Introducing  the notation
\ba
       V_k = C_k / \sqrt{4M_1 \w_1 M_k \w_k}  \,.
\ea
the   Hamiltonian  takes the  form \cite{Karpov00,Antoniou01}
\ba    \label{H}
        H &=&  \w_1 a_1^{\dagger} a_1 + \displaystyle
                    \sum_{k=0}^{\infty} \omega_k a_k^{\dagger} a_k \no
                    &+& \lambda
                    \displaystyle \sum_{k=0}^{\infty} V_k (a_1^{\dagger} + a_1 )
                    (a_k^{\dagger}+a_k) + E_{\rm vac},
\ea
where $ E_{\rm vac}$ is the vacuum energy. The interaction has the
following volume dependence,
\ba  \label{Vk}
        V_k=\left(\frac{2\pi}{L}\right)^{\frac{1}{2}} v_k\,,
\ea
with $v_k$ independent of $L$. In the limit $L \rightarrow \infty$,
the sum over discretized field modes turns into an integral and the
Kronecker delta function  turns into a  Dirac delta function,
\ba \label{LLL}
        \frac{2\pi}{L}\sum_{k} \rightarrow \int dk \quad ,\quad
         \frac{L}{2\pi}\delta_{k,\,k'} \rightarrow \delta(k -k')
         \,.
\ea
Hereafter, whenever we write summations or Kronecker deltas, it is
understood that we take the limit $L\to \infty$ using Eq.
(\ref{LLL}). Also from now on, when we take the limit $L\to \infty$
we will keep the energy density of the field finite. This means that
(with $\bra A \ket = {\rm Tr} (A \rho)$) 
 \ba \label{thldef}
  \bra a_k^\dg a_k \ket \sim L^0, \quad {\rm for}\,\, L \to \infty.
 \ea
This condition is known as the  thermodynamic limit.  We consider
density operators that have diagonal ($\delta$-function)
singularities in field-mode representation  \cite{IP,Petrosky97}. An
example of this class of ensembles is the  equilibrium Gibbs
distribution. For these density operators, 
\ba \sum_{k'} \bra a_k^\dg a_{k'}\ket  \sim  \bra a_k^\dg a_k \ket
\sim O(L^0). \EQN{dsing} \ea 
Diagonal observables are as important as sums of off-diagonal
observables. Due to this property the separation of diagonal and
off-diagonal observables, which we consider below,  is well defined
in  the thermodynamic limit.\footnote{It is possible  to avoid
summations altogether, and use integrals from the beginning
\cite{Laura}. The results are the same.}

The Hamiltonian in \Ep{H} has the form
\ba    \label{H00}
        H = H_0+\lambda V
\ea
where $H_0$ is the unperturbed part describing free motion, and $V$
is the interacting part. Corresponding to this Hamiltonian we have
the Liouville super operator (or ``Liouvillian'')
\ba    \label{L00}
        L_H  &=&  [H,  \,\,\,] \no
                   &=& L_0 + \lam L_V  \,.
\ea

>From  the Liouville equation,
\ba \label{LH0}
        i\frac{\pd}{\pd t} \rho(t) = L_H \rho(t)  \,,
\ea
we obtain the time evolution of averages of observables $W$,
\ba \label{Ave}
      \<W(t) \>=  {\rm Tr} (W \rho(t))  \,.
\ea
We will consider observables depending only on particle operators,
expandable in monomials,
\ba \label{W}
        W={a^\dg_1}^n a_1^m  \,.
\ea
with $m,n\ge 0$ integers.  Then we have 
\ba \label{Ave2}
        \<W(t) \>   =  {\rm Tr} (W \Po \rho(t))
\ea 
where $\Po$ is a linear projection super operator defined by 
\ba   \label{P0}
        & & \Po \left( a_1^{\dg m_1} a_1^{n_1}
            \displaystyle \prod_{k=0}^{\infty}
                {a_k^\dg}^{m_k} a_k^{n_k}\right) \no
                &=& a_1^{\dg m_1} a_1^{n_1} \displaystyle
                \prod_{k}\delta_{m_k,n_k} a_k^{\dg m_k} a_k^{m_k}
                    \,,
\ea 
for $m_i, n_i \ge 0$. This projector singles out products of
creation and annihilation operators with diagonal  field operators.
Every creation operator $a_k^\dagger$ present in the product has to
be paired with the annihilation operator $a_k$.

The projector $  \Po$ commutes with the free Liouvillian $L_0$, 
\ba \label{PLP}
       \Po L_0 =  L_0 \Po.
\ea 
Introducing the complement projector $Q=1-P$ we have $\Po Q = Q \Po
= 0$. Thus under the unperturbed time evolution (with $\lam=0$), any
density operator can be decomposed into two components that evolve
independently 
\ba \label{Subd-}
        \rho = \Po  \rho + Q \rho    \,.
\ea 
Each component remains invariant under the free time-evolution. For
$\lam=0$ we have 
\ba \label{LH1}
        i\frac{\pd}{\pd t} \Po \rho(t) &=& L_0 \Po \rho(t),\nonumber\\
        i\frac{\pd}{\pd t} Q \rho(t) &=& L_0 Q \rho(t).
\ea 
Each component follows its own subdynamics, with closed
time-evolution. This separation allows us to calculate $ \<W(t) \>$
knowing only the $P\rho$ component of $\rho$, without  the
complement $Q\rho$ component.

On the other hand, the interacting Liouvillian (with $\lam\ne 0$)
does not commute with $\Po$. We have 
\ba \label{LH2}
        i\frac{\pd}{\pd t} \Po \rho(t) = \Po L_H \rho(t)  = \Po L_H \Po \rho(t)
         + \Po L_H Q \rho(t) \,.  \no
\ea 
This is no  longer a closed equation for the component $\Po
\rho(t)$. This is the main problem of non-equilibrium statistical
mechanics. A common approach to deal with this problem is to write a
hierarchy of equations of the BBGKY  type \cite{Balescu}.
Alternatively, one can try to obtain closed non-Markovian equations
(with memory terms), such as the Prigogine-Resibois generalized
master equation \cite{Resib}. As shown by Hu, Paz and Zhang, for
the quantum Brownian oscillator it is indeed possible to obtain the
closed non-Markovian equation (\ref{HPZ92}) for the reduced density
matrix. The non-Markovian character of the equation is manifested in
the time-dependent coefficients.

In the subdynamics approach,  we introduce a new projector $ \Pito$
satisfying the following three conditions
\ba
   {\rm(A)}  &&  \qquad {\Pito}^2 = \Pito,  \no
   {\rm(B)} &&\qquad  \Pito L_H = L_H \Pito \no
  {\rm(C)} && \qquad  \Pito= \Po + \lam  \Pito_1 + \lam^2  \Pito_2 + \cdots\nonumber
\ea
where $ \Pito_n$ are independent of $\lam$. The last condition means
that $\lim_{\lam\to 0}  \Pito=  \Po$ and $\Pito$ is analytic at
$\lam=0$ \footnote{As shown in Ref. \cite{Balescu} condition (A)
actually follows from conditions (B) and (C). In this paper we will
verify all three conditions are satisfied.}.

Using $\Pito$ we can decompose a density operator into two
components that evolve independently:
\ba \label{Subd}
        \rho = \Pito \rho + \Pih \rho
\ea
where $\Pih = 1- \Pito$. Each component obeys a closed equation,
\ba \label{LH}
        i\frac{\pd}{\pd t}  \Pito \rho&=& L_H \Pito \rho, \\
       i\frac{\pd}{\pd t} \Pih  \rho&=& L_H \Pih   \rho   \,.
\ea
Hereafter we will focus on the $\Pito$ component. As we will see,
this component gives the closed Markovian equation describing
quantum Brownian motion. The complement component $\Pih$ gives
memory effects associated with dressing \cite{POP2001} \footnote{In
a more detailed formulation of subdynamics (see \cite{TB}) both
$\Pi$ and $\Pih$ are further decomposed into a sum of orthogonal
projectors $\Pi = \sum_\nu \Pi^{(\nu)}$ and $\Pih = \sum_{ \nu}
\Pih^{({\nu})}$. Each sub-component gives a closed Markovian
equation. However, sums of these projectors can give a non-Markovian
equation, as is the case for $\Pih$.}.

We  focus on the equation
\ba \label{KE2}
             i\frac{\pd}{\pd t} {\rm Tr}  [W \Pito \rho] =  {\rm Tr}  [W L_H \Pito \rho]
\ea

In the following, we will derive an explicit form of this equation.
Using the property
\ba \label{trAB}
       {\rm Tr}  [A\cdot S \rho] = {\rm Tr}  \left[ (S^\dg A^\dg)^\dg
            \rho \right] \,.
\ea
where $S$ is a super operator and $A$ is an arbitrary operator, we
have
\ba \label{KE}
             i\frac{\pd}{\pd t} {\rm Tr}  [W \Pito \rho] ={\rm Tr} \left[(L_H {\Pito}^\dg W^\dg )^\dg
                 \rho\right].
\ea
To obtain the kinetic equation, we need to calculate the quantity
$L_H {\Pito}^\dg W^\dg$ with $W={a^\dg_1}^n a_1^{m}$. This will be
done in the following sections.

We note that (see \Ep{Ave2})
\ba \label{KE2'}
           {\rm Tr}  [W L_H \Pito \rho] = {\rm Tr}  [W\Po L_H \Pito \rho ]  \,.
\ea
As shown in Refs. \cite{Balescu,Petrosky97}, we have
\ba \label{LH3}
      \Po L_H \Pito \rho= \thetao \Pito \rho \,,
\ea
where $ \thetao$ is a ``collision''  super operator satisfying the
relation $[\thetao ,\Po]=0$. We get an exact, closed Markovian
equation
\ba \label{LH3'}
        i\frac{\pd}{\pd t}  \Po \Pito \rho=   \thetao \Po\Pito \rho
\ea
for the component $\Po \Pito \rho$. This will be verified through
the direct calculation of \Ep{KE2}.

In the construction of $\Pito$ we will consider two cases discussed
next: integrable case and non-integrable case. As we will show the
Markovian dynamics of Brownian motion occurs in the non-integrable
case. Our approach will be to first obtain $\Pito$ for the
integrable case, and then extend this result to the non-integrable
case.
\section{Integrable  and non-integrable cases}
\label{IntNoInt}

In this section we specify what we mean by integrable and
non-integrable cases.

For the integrable case, the $\Po$ and $\Pito$ projectors can be
related by a similitude transformation, 
\ba \label{PPiU} \Pito = U^{-1} \Po U
      \ea
where $U$ is a {\it time independent} unitary transformation. This is the
same transformation that puts the Hamiltonian in a diagonal form
with no interactions (see Eqs. (\ref{intH}), (\ref{Udef}) below). In
this way the interacting system can be mapped to a non-interacting
system through a unitary transformation.  We call this case
``integrable'' because there exists a one-to-one correspondence
between unperturbed and perturbed invariants of motion. Furthermore,
the perturbed invariants are expandable around $\lam=0$. These
properties were studied by Poincar\'e in the context of celestial
mechanics, so when we speak about integrability, it is in
Poincar\'e's sense \cite{PP88}.

In contrast, for the non-integrable case the interactions cannot be
transformed away through a unitary transformation. There is no more
a one-to-one correspondence between unperturbed and perturbed
invariants. The $\Po$ and $\Pito$ projectors  are now related by a
non-unitary transformation $\Lambda$, 
\ba \label{PPi} \Pito = \Lambda^{-1} \Po \Lambda \,.
      \ea
 As shown in Refs. \cite{Rosen, OPP} the transformation $\Lambda$
is ``star-unitary.'' In this paper we will construct the $\Pito$
projector directly, without using the $\Lambda$ transformation. Let
us  just make a few remarks on this transformation. Rather than
transforming away the interactions, $\Lambda$  takes us from the
original representation in terms of bare particles to a new
representation in terms of dressed particles which obey  stochastic
equations breaking time-symmetry. In this new representation the
effects of noise appear due to the non-distributive character of
$\Lambda$ with respect to multiplication  \cite{Rosen, OPP, KO,
POP}.

For the quantum Brownian oscillator we can have both integrable and
non-integrable cases, depending on the relation between the
frequency of the particle and the frequencies of the field modes.

We assume that the field frequencies $\w_k$ take the values
\ba
      0 \le  \w_{0} \le \w_k < \infty  \,.
 \ea
Here $\w_{0}$ is the lower bound of the spectrum of $\w_k$ for
$k=0$.

The integrable and non-integrable cases correspond, respectively, to
the following two possibilities \cite{Karpov00}:
\ba  \label{Icase}
       (a) \qquad  \w_1<\w_{0}
\ea
\ba  \label{Ncase}
       (b) \qquad  \w_c<\w_1
\ea
where
\ba  \label{wmax}
       \w_c^2 =  \w_{0}^2 + \int_{0}^{\infty}dk\, \frac{4 \w_c \w_k \lambda^2 v^2_k}{\w_k^2 - \w_{0}^2} \,.
\ea
The frequency $\w_c$ is a threshold frequency for $\w_1$, below
which the oscillator goes from damped oscillation to undamped
oscillation. The intermediate case $\w_{0}<\w_1< \w_c$,  gives
undamped oscillations as well. In this case the $\Pito$ and $\Po$
projectors are related  through  a unitary transformation, but this
transformation is not expandable around $\lam=0$. This intermediate
case will not be considered here. Interesting phenomena associated
with this case have been considered in  Refs.
\cite{Suresh,Sterling}.

\section{$\Pi$ in the integrable case}
\label{PiInt}

We consider now the integrable case (a) discussed in section
\ref{IntNoInt}. In this case the particle cannot resonate with the
field modes. The Hamitonian (\ref{H}) can be diagonalized through
the unitary super operator $U$ into the following form \cite{Ant98},
\ba \label{intH}
         H&=& \wb_1 \Ab_1^{\dg} \Ab_1 + \displaystyle
\sum_{k=0}^{\infty} \w_k \Ab_k^{\dg} \Ab_k  \, + {\bar E}_{\rm vac},
\ea
where $\wb_1$ is the renormalized frequency of the particle, ${\bar
E}_{\rm vac}$ is the renormalized vacuum energy, and the operators
$A$ are the renormalized operators of the bare operators $a$. The
$A$ operators are given by the  unitary transformation
\ba   \label{Udef}
        \Ab^\dg_i&=& U^{-1} a_i^\dg   \,, \no
        \Ab_i&=& U^{-1} a_i  \,.
             \ea
We use bars to denote variables in the integrable case. The
transformed operators satisfy the relations
\ba
    \label{EigenLAb}
     &&  L_H \Ab_1^\dg = \wb_1 \Ab_1^\dg,  \quad  L_H \Ab_1 =  -\wb_1 \Ab_1 , \no
      &&  L_H \Ab_k^\dg = \ome_k  \Ab_k^\dg,  \quad  L_H \Ab_k =  -\ome_k
      \Ab_k \,.
\ea
As mentioned in the previous section, we call this case  integrable
because this system follows Poincar\'e's criterion of integrability.
There is a one-to one correspondence between the unperturbed
invariants of motion $a_i^\dg a_i$ and the perturbed invariants
$\Ab_i^\dg \Ab_i$. The perturbed invariants are expandable around
$\lam=0$.

The super operator $U^{-1}$ may be written in the form
\ba
       U^{-1} a = u^{-1} a u
\ea
where $u$ is a unitary operator. Thus we have the distributive
property
\ba \label{distributive}
       U^{-1} ab = [U^{-1} a][U^{-1} b]  \,.
\ea
It follows that the operators $\Ab$ satisfy the commutation
relations
\ba
 \label{Acom}
        [\Ab_i,\,\Ab_j^\dg]=\delta_{ij}
\ea
where $i,j = 1 ,k$.

The  transformed operators are given by the linear relations
\ba   \label{A}
        \Ab^\dg_1&=&\cb_{11}^*a^\dg_1+\db_{11}^* a_1 +\sum_k
                \cb_{1k}^*    a^\dg_k  +\sum_k \db_{1k}^* a_k  , \\
         \Ab^\dg_k&=&  \cb_{k1}^* a^\dg_1+\db_{k1}^* a_1 +\sum_{k'}
                \cb_{kk'}^*    a^\dg_{k'} +\sum_k \db_{kk'}^* a_{k'},
\ea
with the coefficients $\cb$ and $\db$ written in Appendix
\ref{Coefficients}.

The renormalized frequency $\wb_1$ is the solution of the equation
\ba
        G(\wb_1)^{-1} = 0 \,.
\ea
satisfying the condition $\lim_{\lam\to 0} \wb_1 = \w_1$, where $G$
is the Green's function,
\ba \label{Gint}
        G(\w)=\left[ \w_1^2-\w^2-\int_{0}^{\infty}dk\,
                \frac{4 \w_1  \w_k \lambda^2 v^2_k}{\w_k^2-\w^2}\right]^{-1},
\ea
defined here for $\w<\w_0$.
 Using the commutation relations (\ref{Acom}) we  invert (\ref{A}) to obtain
\ba  \label{a}
     a^\dg_1 = \cb_{11}\Ab^\dg_1-\db_{11}^*
            \Ab_1  +\sum_k  \cb_{k1} \Ab^\dg_k  -\sum_k \db_{k1}^* \Ab_k.
\ea
We verify now that in the integrable case, the $\Pibo$ projector is
obtained through the relation
\ba \label{UPU}
        \Pibo= {\Pibo}^\dg  \equiv U^{-1}\Po U \,,
\ea
where we  use a bar to remind that this corresponds to the
integrable case. To prove this, we will check that this expression
satisfies the conditions (A-C) in Sec, \ref{Pisub}.

Condition (A) is satisfied, since $\Po=\Po^2$ itself is a projector.
Condition (B) means that
\ba
       U^{-1}\Po U L_H = L_H  U^{-1}\Po U
\ea
or
\ba \label{c2proof}
       \Po U L_H U^{-1} = U L_H  U^{-1}\Po   \,.
\ea
Using the distributive relation (\ref{distributive}) together with
\Ep{intH} one can show that $U L_H  U^{-1}$ has the same form as the
unperturbed Liouvillian
 $L_0$, which implies Eq. (\ref{c2proof}) is true
\cite{OPP}. Condition (C) is  satisfied, since the super operator
$U$ reduces to the unit super operator when $\lam\to 0$, as can be
seen in the explicit forms of the coefficients in \Ep{A};  see
Appendix \ref{Coefficients}.

The distributive relation (\ref{distributive}) together with Eq.
(\ref{UPU}) also shows that the projector $ \Pibo$ can be defined
through the relation
\ba \label{mP0}
     && \Pibo \left( {\Ab^{\dg m_1}_1} \Ab_1^{n_1} \displaystyle
            \prod_{k=0}^{\infty}
       \Ab_k^{\dg m_k} \Ab_k^{n_k}\right) \no
       &=& {\Ab^{\dg m_1}_1} \Ab_1^{n_1} \displaystyle
        \prod_{k}\delta_{m_k,n_k} \Ab_k^{\dg m_k} \Ab_k^{m_k}
        \,,
\ea
since this is equivalent to Eq. (\ref{P0}).

Note that ${\Pibo}^\dg = {\Pibo}$. Henceforth we write ${\Pibo}^\dg$
anticipating the extension to the non-integrable case (see \Ep{KE}).
\section{Recursive relations}
\label{Recurrence}

Before going to the non-integrable case, we will derive recursive
relations for the $\Pibo$ projector in  the integrable case.
Subsequently, these will be extended to the non-integrable case as a
crucial step in the derivation of our Markovian equation. The
relations are
\ba \label{Pimn+}
   {\Pibo}^\dg ({a^\dg_1}^m a_1^n) &=&
    {\Pibo}^\dg a^\dg_1\cdot  {\Pibo}^\dg ({a^\dg_1}^{m-1}
    a_1^n) \no
      &+& (m-1) \Xb  {\Pibo}^\dg ({a^\dg_1}^{m-2} a_1^n) \no
      &+& n\Yb  {\Pibo}^\dg ({a^\dg_1}^{m-1} a_1^{n-1})
\ea
and
\ba \label{Pimn-}
    {\Pibo}^\dg({a^\dg_1}^m a_1^n)&=&
     {\Pibo}^\dg({a^\dg_1}^{m} a_1^{n-1})\cdot
      {\Pibo}^\dg a_1 \no
      &+& (n-1) \Xb {\Pibo}^\dg({a^\dg_1}^m a_1^{n-2}) \no
      &+& m\Yb  {\Pibo}^\dg ({a^\dg_1}^{m-1} a_1^{n-1})  \,,
\ea
where
\ba
 \label{Xbdef}
\Xb =  -\sum_k \cb_{k1}\db_{k1}^* \{ \Ab_k, \Ab^\dg_k\}   \,, \ea
\ba \label{Ybdef} \Yb =  \sum_k |\cb_{k1}|^2  \Ab^\dg_k \Ab_k +
|\db_{k1}|^2  \Ab_k \Ab^\dg_k \ea
and  \{ \} is the anti-commutator bracket.

In the rest of this section we present the proof of these relations.
To facilitate our construction, we define two new operators, i.e.,
\ba \label{Bbar}
  \Bb^\dg_1\equiv {\Pibo}^\dg a^\dg_1
        &=& \cb_{11}\Ab^\dg_1-\db_{11}^* \Ab_1   \,
\ea
(see \Ep{a}) and
\ba \label{Dbar}
   \Db^\dg_k\equiv \cb_{k1} \Ab^\dg_k - \db_{k1}^* \Ab_k
\ea
with their Hermitian conjugates $\Bb_1$ and $\Db_k$.

We have
\ba \label{aDB}
   a_1^\dg = \Bb_1^\dg + \sum_k\Db^\dg_k
\ea
Thus
\ba \label{Pimn}
  &&   {\Pibo}^\dg({a^\dg_1}^m a_1^n)  \\
  &=& {\Pibo}^\dg\left[ (\Bb^\dg_1+ \sum_k \Db^\dg_k)^m ( \Bb_1+ \sum_k \Db_k)^n \right]   \no
          &=&  {\Pibo}^\dg\left[ \Bb^\dg_1 (\Bb^\dg_1+ \sum_k \Db^\dg_k)^{m-1} ( \Bb_1+ \sum_k \Db_k)^n \right] \no
          &+& {\Pibo}^\dg\left[\sum_k
          \Db^\dg_k (\Bb^\dg_1  + \sum_k \Db^\dg_k)^{m-1} ( \Bb_1+ \sum_k \Db_k)^n \right]. \nonumber
\ea
Using the projection property (\ref{mP0}) of  $\bar{\Pi}$, the first
term of last expression equals
\ba \label{1st}
    &&  {\Pibo}^\dg\left[ \Bb^\dg_1 (\Bb^\dg_1+ \sum_k \Db^\dg_k)^{m-1} ( \Bb_1+ \sum_k \Db_k)^n \right]\no
      &=& \Bb^\dg_1 \cdot  {\Pibo}^\dg\left[ (\Bb^\dg_1
      + \sum_k  \Db^\dg_k)^{m-1} ( \Bb_1+ \sum_k \Db_k)^n \right] \no
          &=& {\Pibo}^\dg a^\dg_1 \cdot  {\Pibo}^\dg
          ({a^\dg_1}^{m-1}a_1^n)   \,.
\ea
Now consider the second term in the right hand side of (\ref{Pimn}).
Expanding this in binomial series we have
\ba
 \label{B6}
 &&  {\Pibo}^\dg \left[ \sum_k \Db^\dg_k \cdot
(\Bb^\dg_1+ \sum_k
          \Db^\dg_k)^{m-1} ( \Bb_1+ \sum_k \Db_k)^n
          \right]         \no
       &=& {\Pibo}^\dg \left[ {\sum_k \Db^\dg_k } \cdot
          \sum_{l=0}^{m-1} C^{m-1}_l
          \Bb^{\dg m-1-l}_1 \left( \sum_{k'}
          \Db^\dg_{k'} \right)^l  \right.\no
           && \left.
          \times\sum_{l'=0}^{n} C^n_{l'}
          \Bb^{n-l'}_1\left(\sum_{k''}
          \Db_{k''}\right)^{l'}\right]   \,.
\ea
where $C^m_l = m!/[(m-l)! l!]$. We have shifted the $\Db$ freely
among $\Bb$ since the $\Db$ operators commute with the $\Bb$
operators (see (\ref{Acom})).

Due to $ {\Pibo}^\dg$, in order to produce a non-trivial projection
the term $\sum_k \Db^\dg_k$ at the beginning of the product must
pair up with one of the $\sum_{k'}\Db^\dg_{k'}$ with $l$ possible
ways or one of the $\sum_{k''} \Db_{k''}$ with $l'$ possible ways.
Checking the volume dependence, we can neglect simultaneous pairings
of $\sum_k \Db^\dg_k$ with three or more $ \Db_k$ or $\Db^\dg_k$,
because such terms will  be of order $O(1/L)$ in the thermodynamic
limit, and therefore can be dropped in comparison with other more
dominant terms. For example,
\ba       \label{O1L}
         {\Pibo}^\dg \left[ \sum_k
            (\Db_k^\dg)^2 (\sum_{k'}
        \Db^\dg_{k'})^2  \right] = \sum_k |\cb_{k1}|^4{\Ab^{\dg 2}_k} {\Ab_k}^2
        \sim   O(1/L) \,, \no
\ea
  where we have used $|\cb_{k1}|^4 \sim |V_k|^4 \sim O(1/L^2)$ and the fact that $\<{\Ab^{\dg 2} }_k {\Ab_k^2 } \> \sim O(L^0) $  in the thermodynamic limit. This last relation follows from  \Ep{thldef}, together with
\ba\label{Akak}
 \bra \At_k^\dg \At_k\ket  = \bra a^\dg_k a_k \ket + O(1/\sqrt{L}).
\ea 
which is due to the volume dependence of the interaction $V_k$.
With this consideration, after suitable relabeling $ \tilde{l}=l-1$
and $\tilde{l'}=l'-1$, \Ep{B6} becomes
\ba \label{2nd}
   &&  {\Pibo}^\dg \left[(m-1)\left({\sum_k \Db^{\dg 2}_k} \right) \sum_{\tilde{l}=0}^{m-2}
         C^{m-2}_{\tilde{l}}
          \Bb^{\dg m-\tilde{l}-2}_1  \right. \no
          && \left. \times \left(\sum_{k'}\Db^\dg_{k'}\right)^{\tilde{l}}
           \left( \Bb_1+ \sum_k \Db_k\right)^n \right] \\
    &+&  {\Pibo}^\dg\left[ n \left({\sum_k \Db^\dg_k \Db_k }\right)  \left(\Bb^\dg_1+ \sum_k
          \Db^\dg_k\right)^{m-1} \right. \no
          && \left.  \times \sum_{\tilde{l}'=0}^{n-1} C^{n-1}_{\tilde l} {\Bb_1}^{n-1-\tilde{l}}\left(\sum_{k'}
          \Db^\dg_{k'}\right)^{\tilde{l'}}\right] \no
    &=& (m-1) \Xb  {\Pibo}^\dg \left[ (\Bb^\dg_1+ \sum_k
          \Db^\dg_k)^{m-2}  ( \Bb_1+ \sum_k \Db_k)^n\right] \no
    &+& n \Yb {\Pibo}^\dg \left[ (\Bb^\dg_1+ \sum_k
          \Db^\dg_k)^{m-1} ( \Bb_1+ \sum_k \Db_k)^{n-1} \right]
          \no
    &=& (m-1) \Xb   {\Pibo}^\dg ({a^\dg_1}^{m-2} a_1^{n}) +n
    \Yb  {\Pibo}^\dg ({a^\dg_1}^{m-1} a_1^{n-1})\nonumber
\ea
where
\ba \label{Y+}
      \Xb = {\Pibo}^\dg \left[ \sum_k {\Db^{\dg 2}_k} \right]
\ea
and
\ba \label{Y0}
      \Yb=   {\Pibo}^\dg \left[ \sum_k \Db^\dg_k  \Db_k \right] \,,
      \ea
which give Eqs. (\ref{Xbdef}) and (\ref{Ybdef}) respectively.

With (\ref{1st}) and (\ref{2nd}), we have proven (\ref{Pimn+}).
Relation (\ref{Pimn-}) can be proven in a similar way.

\section{$\Pi$ in the non-integrable case}
\label{PiNoInt}

In the non-integrable  case (b) discussed in section \ref{IntNoInt},
the particle frequency  $\w_1$ can resonate with the frequencies
$\w_k$ of the field. There is no self-adjoint perturbed invariant
$A_1^\dg A_1$ corresponding to the unperturbed invariant $a_1^\dg
a_1$, which is expandable around $\lam=0$.  This could be expected,
because now the particle is a damped oscillator. Damping comes from
resonant emission of field modes. Due to  damping there is no
invariant of the form $A_1^\dg A_1$ \cite{OPP}.  The Hamiltonian can
now be written as \cite{PPT, Karpov00}
\ba   \label{Hr}
        H=\sum_{k=0}^{\infty}
                \w_k \At^\dg_k \tilde{A}_k  \, + {\tilde E}_{\rm vac} \,, \ea
where $\At_k, \At^\dg_k$ are renormalized annihilation and creation
operators of the field and ${\tilde E}_{\rm vac}$ is the
renormalized vacuum energy in the non-integrable case. As in
scattering theory we can choose either ``in'' or ``out'' operators
\cite{Antoniou01}. Hereafter we will use ``out'' operators. As we
will see,  from the ``out'' operators we will obtain  damping for
$t>0$ in the Heisenberg picture. The explicit form of the operator
$\At_k $   is 
\ba\label{Akdef} \At_k &=&  \ct_{k1}a_1+\dt_{k1}a_1^\dg +\sum_{k'}
                \ct_{kk'}    a_{k'} +\sum_{k'} \dt_{kk'} a_{k'}^\dg\no
\ea 
with the coefficients given in Appendix \ref{Coefficients}. These
coefficients are  proportional to Green's function $G^+(\w_k)$ where
\ba \label{Gz}
        G^\pm(z)=\left[ \w_1^2-z^2-\int_{0}^{\infty}dk\,
                \frac{4 \w_1  \w_k \lambda^2 v^2_k}{(\w_k^2-z^2)_\pm}\right]^{-1}
                \,,
\ea
for general complex argument $z$ with ${\rm Re}(z) > \w_0$. The $+$
($-$) sign means the function is analytically continued from  the
upper (lower) sheet of $z$. The function $G^+(z)$ has a pole on the
``second sheet,'' obtained by analytic continuation form the upper
to the lower half plane of $z$ across the branch cut on the positive
real axis. Denoting this pole as 
\ba
  z_1 \equiv \wt_1 - i\gam
 \ea
(with $\gam>0$) we have  $G^+(z_1)^{-1}=0$. This pole reduces to
$\ome_1$ when $\lam \to 0$.

 By extracting the residue at this pole in \Ep{Hr} we obtain the  complex spectral representation  (see \cite{Antoniou01})
\ba   \label{Hc}
        H= z_1 A^\dg_1 \tilde{A}_1 +\sum_{k=0}^{\infty}
                \w_k \At^\dg_k A_k  \, + {\tilde E}_{\rm vac} , \ea
where 
\ba\label{AntAk}
 A_k = \At_k \left[1 + 2\pi i (\w_k - z_1) \delta_c(\w_k-z_1)\right],
 \ea
$\delta_c$ is the complex $\delta$-function, and 
\ba
  z_1 A^\dg_1 \tilde{A}_1 &=& -{\rm Residue}
                (\w_k \At^\dg_k \tilde{A}_k)_{\w_k=z_1}  \\
 &=& -\sum_{k=0}^{\infty}
                \w_k \At^\dg_k \At_k  2\pi i (\w_k - z_1) \delta_c(\w_k-z_1) \nonumber
\ea 
To evaluate the complex-delta function we first go to the continuous
limit so the summations go to integrals. Then we deform the
integration path to a small contour surrounding the pole $z_1$.

By separating the residue at the pole $z_1$ we obtain  the  particle
operators $A_1^\dg$, $\At_1$ in the complex spectral representation
of the Hamiltonian.  In this way we obtain a closer correspondence
between the integrable and non-integrable cases. Note that the
complex-delta function is a generalized function.

In terms of the non-unitary transformation $\Lambda$ mentioned in
Sec. \ref{IntNoInt} we have
\ba   \label{Lamdef}
        A^\dg_i = \Lambda^{-1} a_i^\dg, \qquad  \At^\dg_i =\Lambda^{\dg} a_i^\dg   \,, \no
        A_i =  \Lambda^{-1} a_i,  \qquad \At_i = \Lambda^{\dg} a_i.
             \ea

The explicit forms of the new operators in \Ep{Hc} are
\ba \label{Ac}
        A^\dg_1  &=&  c_{11}^*a^\dg_1+d_{11}^* a_1 +\sum_k c_{1k}^*
             a^\dg_k  +\sum_k d_{1k}^* a_k   \,, \\
        \At_1 &=&c_{11}^*a_1+d_{11}^* a^\dg_1 +\sum_k c_{1k}^*
             a_k +\sum_k d_{1k}^* a^\dg_k    \,,\no
         A_k &=&  c_{k1} a_1+d_{k1} a_1^\dg +\sum_{k'}
                c_{kk'}    a_{k'} +\sum_{k'} d_{kk'} a_{k'}^\dg\
              \,,  \nonumber
\ea
with the coefficients  presented in Appendix \ref{Coefficients}. The
transformed operators satisfy the relations
\ba \label{LA}
      L_H\At^\dg_1&= z^*_1 \At^\dg_1,    \qquad L_H\At_1
      &=-z_1\At_1 \,, \no
       L_HA^\dg_1&= z_1 A^\dg_1,    \qquad L_H A_1
      &=-z_1^* A_1 \,, \no
      L_H\At^\dg_k&= \w_k \At^\dg_k,    \qquad
      L_H\At_k &= -\w_k \At_k     \,.
\ea

 Due to the complex-delta function,  the operators in Eq. (\ref{Ac}) do not preserve the Hilbert space. For example one can show that (see Appendix \ref{app:A1A1})
\ba\label{noHil}
 [\At_1,\At_1^\dg] = 0.
\ea 
and similarly $[A_1,A_1^\dg]=0$.

 Provided the test functions for integration do not contain singularities at $\w_k = z_1$ or $\w_k = z_1^*$, the new
set of operators  obey the  commutation relations \cite{Antoniou01}
\ba \label{crel}
      &&   \left[\At_1,A_1^\dg\right]= 1   \,,\\
      && \left[\At_k,\At_{k'}^\dg\right] = \left[\At_k,A_{k'}^\dg\right]= \left[A_k,A_{k'}^\dg\right]= \del_{k,k'}  \,.\nonumber
 \ea
Other commutators are zero. If the test functions contain
singularities, then we need a careful consideration \cite{PPT}. Two
examples are presented in Appendix \ref{app:A1A1}.  Hereafter we
assume the density operator $\rho$ gives no such singularities.

From \Ep{Akdef} we have
\ba\label{a1Ak}
        a_1^\dg = \sum_k \Dt^\dg_k
\ea 
where
\ba\label{Dktdef}
        \Dt^\dg_k =  \ct_{k1}\At_k^\dg - \dt_{k1}^* \At_k
\,.
\ea
Separating the poles at $\ome_k=z_1, z_1^*$ from \Ep{a1Ak}  we get
\ba  \label{anoint}
     a_1^\dg = \Bt^\dg_1 + \sum_k D^\dg_k
\ea
where
\ba \label{Bt}
       \Bt^\dg_1 = c_{11} \At^\dg_1 - d_{11}^* \At_1,
\ea
\ba \label{Dt}
      D^\dg_k = \ct_{k1} A^\dg_k - \dt_{k1}^* A_k
       \,,
\ea

Note  that from \Ep{a1Ak} we can calculate the exact time evolution
of $a_1^\dg$ as 
\ba\label{a1tev} e^{i L_H t} a_1^\dg   = \sum_k \left(\ct_{k1} e^{i
\w_k t} \At_k^\dg - \dt_{k1}^* e^{-i \w_k t} \At_k\right)  \,. \ea
From \Ep{a1tev} we can calculate the exact time evolution of any
observable associated with the particle, for example its energy. Our
goal though is to extract  the ``kinetic'' part of the time
evolution, which follows a closed, exact Markovian dynamics. This is
why we  introduce the projector $\Pito$ (or $\Pito^\dg$).

We will calculate the explicit form of the $\Pito^\dg$ projector
acting on products of creation and annihilation operators by
extending the results of the integrable case. As in the integrable
case, the projection $\Pito^\dg (a_1^{\dg m} a_1^n)$ should keep
terms where  the creation operators $\At_k^\dg$ are paired with the
destruction operators  $\At_k$. At the same time,  $\Pito^\dg$
should leave intact particle operators $\At_1^\dg$ and $\At_1$. As
we have seen, the latter are  residues at the poles  $\ome_k=z_1^*,
z_1$.

To define $\Pito^\dg$ we start by  writing  (see \Ep{a1Ak}) 
\ba \label{Pitomn0}
 && a_1^{\dg m} a_1^n \\
 &=&    \sum_{k_1\cdots k_{m+n}}
 \Dt^\dg_{k_1}  \cdots \Dt^\dg_{k_m}  \Dt_{k_{m+1}}  \cdots \Dt_{k_{m+n}} \nonumber
 \ea
We decompose this into a sum of all possible pairings. For example
for $m=n=2$ we have 
\ba \label{Pitomn1}
 && a_1^{\dg 2} a_1^2
=    {\sum_{k_1\cdots k_4}}'
 \Dt^\dg_{k_1}   \Dt^\dg_{k_2}  \Dt_{k_3}  \Dt_{k_4} \\
&+& {\sum_{k_1,k_3,k_4}} ' \Dt^\dg_{k_1}   \Dt^\dg_{k_1}  \Dt_{k_3}
\Dt_{k_4} + {\sum_{k_1,k_2,k_4} }' \Dt^\dg_{k_1}   \Dt^\dg_{k_2}
\Dt_{k_2}  \Dt_{k_4} +  \no &+& \cdots  + {\sum_{k_1,k_2}} '
\Dt^\dg_{k_1}   \Dt^\dg_{k_2}  \Dt_{k_1}  \Dt_{k_2} + \cdots
\nonumber
 \ea
where the prime in the summations means that no summation variables
are equal. Once we have done this separation we can extract the
poles of the unmatched operators using \Ep{anoint}. For example we
have 
\ba \label{Pitomn2}
 &&  {\sum_{k_1,k_2,k_4}} '
\Dt^\dg_{k_1}   \Dt^\dg_{k_2}  \Dt_{k_2}  \Dt_{k_4} \\
&=&{\sum_{k_2}} ' (\Bt^\dg_1 + \sum_{k_1} D^\dg_{k_1})
\Dt^\dg_{k_2}  \Dt_{k_2} (\Bt_1 + \sum_{k_4} D_{k_4})   \,.\nonumber
 \ea
To get the $\Pito^\dg$ projection, we simply drop the unmatched
field operators.  Thus we have 
\ba \label{Pitomn3}
  \Pito^\dg {\sum_{k_1,k_2,k_4}} '
\Dt^\dg_{k_1}   \Dt^\dg_{k_2}  \Dt_{k_2}  \Dt_{k_4} =\sum_{k_2}
\Bt^\dg_1   \Dt^\dg_{k_2}  \Dt_{k_2} \Bt_1  \,.
 \ea
Since the operators with different $k_i$ in the left hand side of
\Ep{Pitomn2} commute, the operators with different index  in the
right hand side of \Ep{Pitomn3} also commute.

In general we can write 
\ba \label{Pitomn}
 \Pito^\dg (a_1^{\dg m} a_1^n) =   \Pito^\dg \left[(\Bt^\dg_1 + \sum_k D^\dg_k)^m (\Bt_1 + \sum_k D_k)^n \right]
\ea 
where the projection in \Ep{Pitomn}   is defined as follows
 \ba \label{mP0ni}
     && {\Pito}^\dg  \left( {\At^{\dg m_1}_1} \At_1^{n_1} \displaystyle
            \prod_{k=0}^{\infty}
       A_k^{\dg m_k} A_k^{n_k}\right) \no
       &=& {\At^{\dg m_1}_1} \At_1^{n_1} \displaystyle
        \prod_{k}\delta_{m_k,n_k} A_k^{\dg m_k} A_k^{m_k},
\ea
which corresponds to \Ep{mP0} in the integrable case.

In the Heisenberg picture, \Ep{mP0ni} decays for $t>0$ and  $m_1,
n_1>0$.  If we had started with the ``in'' operators we would obtain
decay for $t<0$. For $m_1=n_1=0$, \Ep{mP0ni} remains invariant.

>From \Ep{Hr} we see that $\Pi^\dg H = H$, hence 
\ba\label{PiInv}
 {\rm Tr} (H \rho) = {\rm Tr} (H \Pito \rho)
\ea 
which shows that $\rho_{\rm eq} = \Pito \rho_{\rm eq}$ for the
equilibrium distribution.
 Similar to \Ep{PiInv}, for any invariant observable $I$ we have
\ba\label{PiInv2}
 {\rm Tr} (I \rho) = {\rm Tr} (I \Pito \rho) \,.
\ea 
Therefore the state $\Pito \rho$ contains the total  energy and
probability  of $\rho$. For non-invariant observables, $\Pito \rho$
extracts the purely exponential terms of the time evolution, which
are associated with the complex energies $z_1, z_1^*$. In this way,
$\Pito \rho$ gives the Markovian dynamics of approach to
equilibrium. The complement component $\Pih \rho$ extracts
non-exponential terms that give memory effects. This non-Markovian
component contains no net energy or probability.

Following the same steps as in the integrable case  we obtain
recursive relations corresponding to Eqs. (\ref{Pimn+}) and
(\ref{Pimn-}),
\ba \label{Pim+1n}
     {\Pito}^\dg(a_1^{\dg m+1} a_1^n)&=&
    {\Pito}^\dg a^\dg_1 {\Pito}^\dg(a_1^{\dg m}
    a_1^n)+ m \Xt {\Pito}^\dg(a_1^{\dg m-1} a_1^n) \no
      &+& n \Yt {\Pito}^\dg(a_1^{\dg m} a_1^{n-1})    \,,
\ea
\ba \label{Pitmn-}
    {\Pito}^\dg(a^{\dg m}_1 a_1^{n+1})&=&
     {\Pito}^\dg(a^{\dg m}_1 a_1^{n})
     {\Pito}^\dg  a_1 + n \Xt  {\Pito}^\dg(a^{\dg m} a_1^{n-1}) \no
      &+& m\Yt  {\Pito}^\dg( a^{\dg m-1}_1 a_1^{n})
\ea
where
\ba X =  -\sum_k \ct_{k1}\dt_{k1}^* \{ A_k, A_k^\dg\}   \,, \ea
\ba Y =  \sum_k |\ct_{k1}|^2  A_k^\dg A_k + |\dt_{k1}|^2  A_k
A_k^\dg \ea

The $\Pito$ super operator we introduced satisfies all the
conditions of Sec. \ref{Pisub}.   The validity of  condition (A)
is self-evident from the identification of ${\Pito}^\dg$ with the
projector in (\ref{mP0ni}).

The second condition (B) can be verified from the relations 
\ba
 L_H  {\cal A} = z  {\cal A} , \quad \Pito^\dg  {\cal A}   = \xi   {\cal A}  \,,
  \ea
with 
\ba
  {\cal A}   &=&   \At_1^{\dg m_1} \At_1^{n_1} \prod_k   A_k^{\dg m_k} A_k^{n_k} \,, \no
 z&=&   m_1 z_1^* - n_1 z_1 +  \sum_k (m_k - n_k) \w_k \,, \no
 \xi &=& 0,1   \,.
\ea 
This implies that ${\rm Tr} ( {\cal A} [L_H, \Pito] \rho) = 0$ for
all observables $ {\cal A} $ and their linear combinations.

The third condition (C) is verified in Appendix \ref{Analyticity}
using the recursive relations for $\Pito^\dg$.

\section{Exact Markovian Kinetic Equation}
\label{Exact}

With all the previous preparations, we are ready now to derive the
explicit form of the Markovian equation (\ref{KE}). As we saw in
Section \ref{Pisub}, in order to obtain this equation we need to
evaluate $L_H{\Pito}^\dg W^\dg = L_H{\Pito}^\dg  (a_1^{\dg m}
a^n_1)$.

>From the recursive relation (\ref{Pim+1n}) we get
\ba  \label{LHPiaa}
       &&  L_H{\Pito}^\dg(a^{\dg m}_1 a^n_1) \\
        &=& L_H\Bt^\dg_1\cdot
          {\Pito}^\dg(a^{\dg m-1}_1 a^n_1)+\Bt^\dg_1\cdot L_H {\Pito}^\dg(a^{\dg m-1}_1
          a^n_1)\no
          &+& (m-1) \Xt L_H {\Pito}^\dg(a^{\dg m-2}_1 a^n_1)
          +n \Yt L_H{\Pito}^\dg(a^{\dg m-1}_1 a^{n-1}_1)     \,, \nonumber
\ea
where $L_H \Xt=0=L_H \Yt $ since $\Xt$ and $\Yt$ are diagonal in the
transformed  creation-annihilation operators of the field.

Writing
\ba \label{AinB}
         \At^\dg_1=\frac{1}{\Delta}(c_{11}^* \Bt^\dg_1+d_{11}^*\Bt_1 ), \quad
         \At_1=\frac{1}{\Delta}(c_{11}\Bt_1+d_{11}\Bt^\dg_1 ), \no
\ea
where
\ba \label{Delta}
        \Delta =|c_{11}|^2-|d_{11}|^2=|N|^2\frac{\wt_1}{\w_1}   \,.
\ea
and using (\ref{LA}), we find that
\ba
        L_H\Bt^\dg_1&=&\alpha^* \Bt^\dg_1+\beta \Bt_1,\no
          L_H\Bt_1&=&-\beta^* \Bt^\dg_1-\alpha \Bt_1,
\ea
where
\ba \label{albt}
      \alpha&=&
      \frac{\wt^2_1+\gam^2+\w^2_1}{2\w_1}-i\gam
      \,,
      \no
      \beta&=& \frac{\wt^2_1+\gam^2-\w^2_1}{2\w_1}-i\gam  \,.
\ea

>From \Ep{LHPiaa} we then infer that
\ba \label{Lmn}
       && L_H {\Pito}^\dg(a^{\dg m}_1 a^n_1) \\
              &=&
         (m\al^*-n\al) {\Pito}^\dg(a^{\dg m}_1 a^n_1)     \no
          &+& m\bt{\Pito}^\dg(a^{\dg m-1}_1 a^{n+1}_1) -n\bt^*{\Pito}^\dg(a^{\dg m+1}_1
          a^{n-1}_1)\no
          &-&mn\left[(\al^*-\al)\Yt +(\bt-\bt^*) \Xt \right]{\Pito}^\dg(a^{\dg m-1}_1
          a^{n-1}_1)\no
     &-& m(m-1)\left[\al^* \Xt +\bt \Yt  \right] {\Pito}^\dg(a^{\dg m-2}_1
          a^{n}_1)\no
          &+&n(n-1) \left[\al \Xt +\bt^* \Yt  \right] {\Pito}^\dg(a^{\dg m}_1
          a^{n-2}_1) \,. \nonumber
\ea

The consistency between Eqs. (\ref{LHPiaa}) and (\ref{Lmn}) can be
proven by first inserting the recursive relations for
${\Pito}^\dg(a^{\dg m}_1 a^n_1)$, i.e., Eqs. (\ref{Pim+1n}) and
(\ref{Pitmn-}), together with  Eq. (\ref{Lmn}) into the right hand
side of Eq. (\ref{LHPiaa}). Then we verify that the coefficients of
${\Pito}^\dg(a^{\dg m}_1 a^n_1)$ for all values of $m$ and $n$ on
the right hand side of Eq.  (\ref{LHPiaa}) are identical to the
corresponding coefficients in Eq. (\ref{Lmn}).

Now we come to our kinetic equation. We have the relations
\ba \label{PitY}
           {\Pito}^\dg(a^{\dg m}_1
                a^n_1\cdot \Yt )&=& \Yt  \cdot{\Pito}^\dg(a^{\dg m}_1
                a^n_1), \no
         {\Pito}^\dg(a^{\dg m}_1
                a^n_1\cdot \Xt )&=& \Xt  \cdot{\Pito}^\dg(a^{\dg m}_1
                a^n_1)\,.
\ea
because $\Yt$ and $\Xt$  are diagonal in the field operators, and
${\Pito}^\dg$ is a projection of the diagonal component  of the
field operators.

Using (\ref{PitY}), we write (\ref{Lmn}) as
\ba \label{LPinm}
           L_H {\Pito}^\dg(a^{\dg m}_1 a^n_1) = {\Pito}^\dg \Xi
    \ea
where
\ba \label{LPinm'}
           \Xi
           &=&
          (m\al^*-n\al)(a^{\dg m}_1 a^n_1)     \no
         &+&   m\bt (a^{\dg m-1}_1 a^{n+1}_1) -n\bt^*(a^{\dg m+1}_1
          a^{n-1}_1)\no
          &-&  mn\left[(\al^*-\al)\Yt +(\bt-\bt^*) \Xt \right] (a^{\dg m-1}_1
          a^{n-1}_1)\no
           &-&  m(m-1)\left[\al^* \Xt +\bt \Yt  \right] (a^{\dg m-2}_1
          a^{n}_1)\no
          &+&  n(n-1) \left[\al \Xt +\bt^* \Yt  \right] (a^{\dg m}_1
          a^{n-2}_1) \,. \nonumber
\ea
Defining 
\ba
 \rhoo = \Pito \rho \,,
\ea 
the kinetic equation (\ref{KE}) becomes \ba
  i\frac{\pd}{\pd t}
        {\rm Tr} [W \rhoo] =  {\rm Tr} [\Xi^\dg \rhoo]  \,.
\ea 

Using the identities for the trace in Appendix \ref{AppC}, and
defining $\al_R = {\rm Re} (\al)$, $\bt_R = {\rm Re} (\bt)$, we
obtain
\begin{widetext}
\ba
     \label{master}
     &&  i\frac{\pd}{\pd t}
        {\rm Tr} [W \rhoo] = {\rm Tr}  \bigg[ W \cdot \bigg\{ \al_R\left(
            [ a^\dg_1 a_1, \rhoo]  \right)         \\
               &-&i\gam( \Yt -\Xt+\frac{1}{2}) \left([
         a_1\rhoo,a^\dg_1]+[a_1,\rhoo a^\dg_1]
            +[a^\dg_1\rhoo,a_1]+[a^\dg_1,\rhoo a_1]\right)\no
               &+&i\frac{\gam}{2} \left([
         a_1\rhoo,a^\dg_1]+[a_1,\rhoo a^\dg_1]
            -[a^\dg_1\rhoo,a_1]-[a^\dg_1,\rhoo a_1]\right)\no
               &+& \bt_R\left([
         a^\dg_1,\rhoo a^\dg_1]-[a_1\rhoo,a_1]\right)
           + i\gam\left([a^\dg_1,\rhoo a^\dg_1]+[a_1\rhoo,a_1]\right)\no
              &+&( \al_R\Xt +  \bt_R \Yt  )\left([a^\dg_1\rhoo,a^\dg_1]+[a^\dg_1,\rhoo a^\dg_1]
              -[a_1\rhoo,a_1]-[a_1,\rhoo a_1]\right)\no
               &+& i\gam (\Yt -\Xt)\left([a^\dg_1\rhoo,a^\dg_1]+[a^\dg_1,\rhoo a^\dg_1]
             +[a_1\rhoo,a_1]+[a_1,\rhoo a_1]\right)
             \bigg\}
             \bigg]
             \,.\nonumber
\ea
\end{widetext}
In this form we can already see an interesting property of this
equation: it is closed for the component  $ \Po \rhoo$, as mentioned
after \Ep{LH3}. To see this, note that  ${\rm Tr} [W \rhoo] = {\rm
Tr} [W P \rhoo] $. Moreover we have $[P,Y] \sim [P,X]\sim
O(1/\sqrt{L})$. Thus in the thermodynamic limit we can move $P$ past
$X$ and $Y$. By its definition, we can also move $P$ past $a_1,
a_1^\dg$.  So in \Ep{master}  we can  replace $\rhoo$ by $\Po\rhoo$.
This shows that this is a closed equation for $\Po\rhoo$.

The equation takes a simpler form if we assume that initially the
density matrix is factored into particle and field components, 
\ba
 \label{rhot0}
 P  \rhoo(0) = \rhoo_1(0)\times \prod_k \rhoo_k(0)  \,.
\ea 
To see this, we write \Ep{master} as 
\ba
  i\frac{\pd}{\pd t}
        {\rm Tr} [W \rhoo]  =  {\rm Tr} [W \theta \rhoo]
\ea 
with the formal solution 
\ba\label{fs}
        {\rm Tr} [W \rhoo(t)]  =  {\rm Tr} [W e^{-i \theta t} \rhoo(0)]  \,.
\ea 
The collision super operator has the following operator dependence
\ba
  \theta = \theta(X, Y, a_1^\dg, a_1)  \,.
\ea 
Following an argument similar to the one above \Ep{O1L} (see also
Ref. \cite{KO}), we may neglect correlations for $Y$ and $X$
products, i.e., 
\ba
 \< Y^n X^m \> = \< Y \>^n  \< X \>^m + O(1/L)
\ea 
where using Eqs. (\ref{Akak}) and (\ref{rhot0}) we have 
\ba
 \< Y \> = {\rm Tr}_F(Y \prod_k \rhoo_k(0)).
\ea
and similarly for $X$.  Thus in \Ep{fs} we can replace $X$ and $Y$
by their initial averages: 
\ba
  \theta = \theta(\< X \>, \< Y \>, a_1^\dg, a_1)  \,.
\ea 
This means that neglecting $O(1/L)$ terms, only the particle
component of the density operator evolves in time, 
\ba
 \label{rhott}
 P  \rhoo(t) = \rhoo_1(t)\times \prod_k \rhoo_k(0)
\ea 
and the field density operator drops out after taking the trace.
Using this result  and defining the reduced density matrix 
\ba
 \rhoo_1 (x_1,x_1') =  \bra x_1| \rhoo_1 |x_1' \ket
\ea 
we  write the kinetic equation in the dimensionless coordinate
representation as (see Appendix \ref{AppD}),
\begin{widetext}
 \ba   && i\frac{\pd}{\pd t} \rhoo_1 \\
        &=&  \bigg\{ -\frac{ \al_R}{2}\left(\frac{\pdi^2}{\pdi x^2_1}
              -\frac{\pdi^2}{\pdi {x'}^2_1}\right)+\frac{\al_R}{2}(x^2_1-{x'}^2_1)\no
    &+&i\gam (\bra \Yt \ket -\bra \Xt \ket +\frac{1}{2})\left[\left( \frac{\pdi}{\pdi {x}_1}
              +\frac{\pdi}{\pdi {x'}_1}\right)^2-(x_1-{x'}_1)^2
              \right] \no
    &+&i\frac{\gam}{2}\left[  \left( \frac{\pdi}{\pdi {x}_1}+\frac{\pdi}{\pdi {x'}_1}\right)(x_1+{x'}_1)
         -(x_1-{x'}_1)\left( \frac{\pdi}{\pdi {x}_1}-\frac{\pdi}{\pdi
         {x'}_1}\right)\right]\no
    &+&\frac{\bt_R}{2}\left[ (x^2_1-{x'}^2_1)+\left(\frac{\pdi^2}{\pdi x^2_1}
          -\frac{\pdi^2}{\pdi {x'}^2_1}\right)+2(x_1-{x'}_1)+\left( \frac{\pdi}{\pdi {x}_1}
          +\frac{\pdi}{\pdi {x'}_1}\right)\right] \no
    &-&i\frac{\gam}{2} \bigg[ (x_1-{x'}_1)^2 + \left( \frac{\pdi}{\pdi {x}_1}
             +\frac{\pdi}{\pdi {x'}_1}\right)^2  \no
  &+&(x_1-{x'}_1)\left( \frac{\pdi}{\pdi {x}_1}
             -\frac{\pdi}{\pdi {x'}_1}\right)
              + \left( \frac{\pdi}{\pdi {x}_1}+\frac{\pdi}{\pdi
              {x'}_1}\right)(x_1+{x'}_1) \bigg]  \no
    &+& ( \al_R\bra \Xt \ket +  \bt_R \bra \Yt \ket )\left[2(x_1-{x'}_1)\left( \frac{\pdi}{\pdi {x}_1}
          +\frac{\pdi}{\pdi {x'}_1}\right)\right] \no
    &+&i\gam (\bra \Xt \ket-\bra \Yt \ket)\left[(x_1-{x'}_1)^2 + \left( \frac{\pdi}{\pdi {x}_1}
             +\frac{\pdi}{\pdi {x'}_1}\right)^2\right] \bigg\} \rhoo_1 \,, \nonumber
\ea
since it is true when averaging over arbitrary polynomial $W$.
Recollecting terms we finally have our exact Markovian equation
\ba
 \label{OurEq}
  && i\frac{\pd}{\pd t} \rhoo_1 \\
        &=& \frac{|z_1|^2}{2\w_1}(x^2_1-{x'}^2_1)\rhoo
           -\frac{\w_1}{2}\left(\frac{\pdi^2}{\pdi x^2_1}
              -\frac{\pdi^2}{\pdi
              {x'}^2_1}\right)\rhoo_1 \no
        &-& i\gam(x_1-{x'}_1)\left(\frac{\pdi}{\pdi x_1}
          -\frac{\pdi}{\pdi {x'}_1}\right)\rhoo_1 -2i\gam
          K(x_1-{x'}_1)^2\rhoo_1\no
        &+& \w_1 J (x_1-{x'}_1)\left(\frac{\pdi}{\pdi x_1}
          +\frac{\pdi}{\pdi {x'}_1}\right)\rhoo_1
          \,, \nonumber
\ea
\end{widetext}
 The coefficients
are given by 
\ba
             K&=&\bra \Yt \ket - \bra \Xt \ket + \frac{1}{2} \,, \no
        J&=&\frac{|z_1|^2}{\w_1^2} J' -K  \,, \no
           J'&=&\bra \Yt \ket + \bra \Xt \ket + \frac{1}{2}  \,.
\ea 
Using the identities \cite{Karpov00} 
\ba
 \sum_k \w_k \left( |\ct_{k1}|^2 + |\dt_{k1}|^2 \right) = \w_1 \,, \quad \sum_k \w_k \ct_{k1}^* \dt_{k1} = 0, \no
\ea 
we get \ba
    K&=&4\int_{0}^\infty dk\,\lam^2v^2_k|G^+(\w_k)|^2\w^2_k \left[n_k
        +\frac{1}{2}\right]  \,, \\
    J'&=&4\int_{0}^\infty dk\,\lam^2v^2_k|G^+(\w_k)|^2\w^2_1 \left[n_k
        +\frac{1}{2}\right] \,, \no
        J &=& 4\int_{0}^\infty dk\,\lam^2v^2_k|G^+(\w_k)|^2(|z_1|^2 -\w^2_k) \left[n_k
        +\frac{1}{2}\right] \,, \nonumber
\ea 
where $n_k= \<a^\dg_k a_k\>$.

Note that for the integrable case, with no particle-field resonance,
we have $\gam=0$ and the damping and diffusion terms in \Ep{OurEq}
vanish. Still,  the last term remains. This corresponds to the
excitation of the particle due to virtual processes.

We also note that the kinetic equation (\ref{master}) is valid for
all field distributions, while \Ep{OurEq}  is valid for field
distributions of the form (\ref{rhot0}), which are more general than
Gibbsian distributions.

\section{Comparison with the Hu-Paz-Zhang Kinetic Equation  }
\label{HPZKE}

In this section we will obtain an expression for the kinetic
equation correct up to $O(\lam^2)$, in the weak coupling limit
$\lam\to 0$. We then compare this to the $\lam^2t$-limit of the HPZ
equation and show that they are identical. The $\lam^2 t$ limit
means that we take $\lam\to 0, t\to\infty$ with $\lam^2 t$ finite.
Physically, this means that we consider times of the order of the
relaxation time $t_{\rm rel} \sim 1/\lam^2$ in a weakly coupled
system.

We start by assuming the harmonic oscillator is interacting with a
bath of field modes that is in thermal equilibrium, and the number
density of the field degrees of freedom $ n_k $ satisfies the Planck
distribution. We then have 
\ba
        n_k +\frac{1}{2}&=& \frac{1}{e^{\bt\hbar \w_k}-1} + \frac{1}{2}
                \no
                &=& \frac{1}{2} {\rm coth}\left( \frac{\bt\hbar \w_k}{2}\right)  \,,
\ea 
where $\bt=1/k_B T$ and $k_B$ is the Boltzmann constant.

In what follows, we are going to use the approximation of the
function $G^\pm(\w_k)$   shown in Appendix \ref{pole}. First we
consider the coefficient $J$, 
\ba \label{Jlam2}
        J = - 4 \int_{0}^\infty dk\, \lam^2 v^2_k |G^+_k|^2(w_k^2- |z_1|^2)  ( n_k +\frac{1}{2}) \,.
\ea 

>From \Ep{G2''} we find that  in the zeroth order approximation
\ba \label{G3}
         \lam^2 v_k^2  |G^+(\ome_k)|^2 (\w_k^2 - \w_1^2)    =  0   \,.
\ea
The next order can be found using the expression (see \Ep{G+k})
\ba \label{G2}
       |G^+(\ome_k)|^2    \simeq \frac{1}{(\w^2_k-\w^2_1)^2+\eps^2}  \,.
\ea
which gives
\ba \label{G4}
         \lam^2 v_k^2  |G^+(\ome_k)|^2 (\w_k^2 - \w_1^2)
 & \simeq &   \lam^2 v_k^2  \frac{\w_k^2 - \w_1^2}{(\w^2_k-\w^2_1)^2+\eps^2} \no
&=&   \lam^2 v_k^2 {\mathcal P}\frac{1}{\w_k^2 - \w_1^2}   \,. \ea
Thus we have 
\ba \label{Jlam3}
        J
           &\simeq& -2 \int_{0}^\infty dk\, \lam^2 v^2_k \, \mathcal{P} \left( \frac{1}{\w^2_k-\w^2_1}\right)
                  {\rm coth}\left( \frac{\bt \hbar \w_k}{2}\right)    \,, \no
\ea 

For the coefficient $K$ we have using (\ref{G2''}) 
\ba      K                &\simeq& \frac{1}{2\w^2_1}\int_{0}^\infty
d\w_k\, \delta(\w_k-\w_1) \w^2_k
             {\rm coth}\left( \frac{\bt \hbar \w_k}{2}\right)    \no
             &=& \frac{1}{2} {\rm coth}\left( \frac{\bt \hbar \w_1}{2}\right)   \,.
\ea 
Therefore, in the $\lam^2$-approximation, the kinetic equation is 
\ba \label{KElam2}
           i\frac{\pd}{\pd t}  \rhoo &=&  \bigg[
                            -\frac{\w_1}{2}\left(\frac{\pdi^2}{\pdi x^2_1}
              -\frac{\pdi^2}{\pdi{x'}^2_1}\right)  + \frac{\wt^2_1}{2\w_1} (x^2_1-{x'}^2_1)  \no
               &-& i\gam (x_1-x'_1)\left(\frac{\pdi}{\pdi x_1}
              -\frac{\pdi}{\pdi{x'}_1}\right)  \no
              &-& i\gam
                {\rm coth}\left( \frac{\bt \hbar \w_1}{2}\right)
                (x_1-x'_1)^2   \no
                &-& \mathcal{P} \int_{0}^\infty dk\, \frac{2\w_1 \lam^2 v^2_k }{\w^2_k-\w^2_1}
                {\rm coth}\left( \frac{\bt \hbar \w_k}{2}\right) \no
                &\times& (x_1-x_1') \left(\frac{\pdi}{\pdi x_1}
              + \frac{\pdi}{\pdi{x'}_1}\right)  \bigg]\rhoo    \,.
\ea 
Now we show that this expression is exactly the same as the $\lam^2
t$ limit of the HPZ equation. For weak coupling,  the time dependent
coefficients of the HPZ equation (\ref{HPZ92}) are found to be
\cite{HPZ} 
\ba
       \tilde{\Omega}^2 (t) &\simeq& \w_1^2 + \delta\Omega^2(t)   \,, \\
        \delta\Omega^2(t) &\simeq& 2 \int_{0}^{t} ds \, \eta(s) \cos (\w_1
        s)  \,, \\
        \Gamma(t) &\simeq& - \frac{1}{\w_1} \int_{0}^{t} ds \, \eta(s) \sin (\w_1
        s)  \,, \\
        \Gamma(t) h(t) &\simeq&  \int_{0}^{t} ds \, \nu(s) \cos (\w_1
        s)  \,, \\
        \Gamma(t) f(t) &\simeq&  \frac{1}{\w_1} \int_{0}^{t} ds \, \nu(s) \sin (\w_1
        s)  \,,
 \ea
 where
 \ba
        \eta(s) &=& - \int_{0}^{\infty} d\w \, I(\w) \sin (\w_1
        s)  \,, \\
        \nu(s)  &=& \int_{0}^{\infty} d\w \, I(\w) \coth (\frac{\bt \hbar
        \w}{2}) \cos( \w_1 s)   \,, \\
        I(\w) &=& \frac{1}{M_1} \sum_{k=0}^{\infty} \delta(\w -
        \w_k) \frac{\lam^2 C_k^2}{2 M_k \w_k} \no
           &=& 2 \w_1  \sum_{k=0}^{\infty} \delta(\w -
        \w_k) \lam^2 V_k^2  \no
                  &=& 2 \w_1 \lam^2 \vt^2(\w).
\ea 
We note that we have added in an extra factor of $1/M_1$ in the
expression of $I(\w)$ compared to \cite{HPZ} due to the different
definition of the position $x_1$.

In the $\lam^2 t$ limit we then take  $t\to\infty$. We use the trick
in \cite{TB} to evaluate the time integrations. For example we have
\ba
     &&   \int_{0}^{\infty} ds \,  \sin (\w_1 s) \sin (\w s)
        =  {\rm lim}_{\eps \rightarrow 0^+} \frac{-1}{4} \int_{0}^{\infty} ds \no
       &\times& \left[ e^{i(\w_1+\w+i \eps)s} + e^{-i(\w_1+\w - i \eps)s} \right.\no
          &-& \left.  e^{i(\w_1-\w+i \eps)s} - e^{-i(\w_1-\w - i \eps)s}
           \right]     \no
        &=& - \frac{1}{4}  \left[  \frac{-1}{i(\w_1+\w+i\eps)}
        -\frac{1}{-i(\w_1+\w - i \eps)} \right. \no
       &+&\left. \frac{1}{i(\w_1-\w + i
        \eps)}+\frac{1}{-i(\w_1-\w-i\eps)}  \right]    \no
        &=&  - \frac{1}{4i}  \left[ 2 i \pi \delta(\w_1+\w) -2 i
        \pi \delta (\w_1 -\w)  \right]     \no
        &=& \frac{\pi}{2} \delta(\w_1-\w)  \,.
\ea 
The last line is due to the fact that $\w$ can only be positive.

In this way we can  calculate all the  coefficients. Using the
expressions for $\wt_1$ and $\gam$  in Appendix \ref{app:z1} we find
\ba
        \delta\Omega^2(t) &\simeq& -2 \int_{0}^{\infty} d\w \, I(\w)
        {\cal P}\frac{\w}{\w^2-\w_1^2}   \no
        &=& 4 \int_{0}^\infty dk\, {\cal P} \frac{\w_1 \w_k \lam^2
        v^2_k}{\w_1^2-\w_k^2}   \no
        &=& \tilde{\w}_1^2 - \w_1^2  \,,
\ea 
\ba
        \Gamma(t) &\simeq& \frac{1}{\w_1} \int_{0}^{\infty} d\w \, I(\w) \frac{\pi}{2} \delta(\w_1-\w)
        \no
        &=& \frac{\pi}{2\w_1} I (\w_1) =  \pi \lam^2 \vt^2(\w_1)   \no
        &=& \gam  \,,
\ea 
\ba
        \Gamma(t) h(t) &\simeq&   \int_{0}^{\infty} d\w \, I(\w)
        \coth ( \bt \hbar \w/2 )  \frac{\pi}{2} \delta(\w-\w_1)
        \no
        &=& \frac{\pi}{2 } I(\w_1) \coth(\bt \hbar \w_1 /2) \no
        &=& \w_1 \gam \coth(\bt \hbar \w_1 /2)   \,,
\ea 
\ba
         \Gamma(t) f(t) &\simeq& - \lam^2 \int_{0}^{\infty} d\w \, I(\w)
        \coth ( \bt \hbar \w/2 ) {\mathcal P}\frac{1}{\w^2-\w^2_1}     \no
        &=&  -2 \w_1 {\mathcal P} \int_{0}^\infty dk\, \frac{\lam^2
        v^2_k}{ \w^2_k - \w^2_1 } \coth(\bt \hbar \w_k /2)   \,.\no
\ea
Upon comparison of the $\lam^2 t$-approximation of the HPZ equation
with the $\lam^2$-approximation of our exact Markovian kinetic
equation (\ref{KElam2}), we find that they are identical
\footnote{In Ref. \cite{TB} the operators $P^{(\nu_1)}$ were used to
construct the subdynamics. As a result, the ``anomalous'' diffusion
term in the HPZ equation did not appear in the kinetic equation
derived in Ref. \cite{TB}, because this term belongs to the
``non-privileged'' components.   In order to obtain this term one
has to include  ``non-privileged'' components. In contrast, in the
present paper we use the  operator  $P=\sum_{\nu_1} P^{(\nu_1)}$ and
the anomalous diffusion term is included in the privileged
components. A detailed calculation shows that both the HPZ equation
and our equation are consistent with Eq. (136) of Ref. \cite{TB} in
the one-particle sector. The anomalous diffusion term involves
higher particle sectors, which were not considered in \cite{TB}. }.

Beyond the weak coupling limit, our equation gives the Markovian
dynamics of the quantum Brownian oscillator valid even for strong
coupling, and also valid for any time scale. For $t\to\infty$ the
solution of our equation gives the equilibrium solution of the
complete dynamics. This is so because the equilibrium distribution
is a function of the Hamiltonian, and any function of the
Hamiltonian belongs to the $\Pito$ subspace (see \Ep{PiInv}). The
complement component $\Pih\rho$ gives all the memory effects, which
vanish for $t\to\infty$.

\section{Concluding Remarks}
\label{Conclusions}
The example presented in  this paper shows that irreversible
Markovian dynamics can be regarded as an exact dynamics taking place
in the subspace of density operators $\Pito\rho$, for non-integrable
systems in the sense of Poincar\'e. The breaking of time-symmetry in
the equation 
\ba \label{LH22}
        i\frac{\pd}{\pd t}  \Pito \rho= L_H \Pito \rho.
\ea 
is due to $\Pito$ being non-Hermitian, and appears before we take
the trace over the field. Thus from our point of view of
irreversibility, rather than a consequence of coarse graining,  is a
property of the invariant subspaces of the Liouvillian.
Time-symmetry breaking appears because the construction of $\Pito$
involves generalized creation and annihilation operators ($A_1$,
$\At_1$). These are  eigen-operators of the Liouvillian with complex
eigenvalues ($z_1$, $z_1^*$) either in the lower or upper
half-planes.  In this formulation we add no extra dissipative terms
to the Liouvillian.

The formulation presented in this paper links  stochastic processes
and dynamics in a direct  way.  Once we have a Markovian kinetic
equation we have a stochastic process described by Langevin-type
equations without any memory terms. An interesting question is to
see what is the spectrum of quantum noise associated with such
Langevin equations (see also \cite{Ford}).

In this paper we focused on the $\Pito \rho$ component of the
density matrix. In a sense, this component corresponds to
traditional thermodynamics. From the Markovian kinetic equation we
can  derive a non-equilibrium entropy and the second law of
thermodynamics even for strong coupling. This could be considered in
a subsequent publication.

In contrast to $\Pito \rho$, the complement component $\Pih \rho$
gives ``non-traditional'' thermodynamics including memory effects.
Deviations from thermodynamics in small quantum systems have been
reported in Ref. \cite{Nieu}. It would be interesting to see how the
behavior of  $\Pih \rho$ is related to these deviations, and what
type of non-Markovian equation is obtained for this component.

The model we considered is exactly solvable. For systems with
non-linear interactions we have to use a perturbative approach.  It
is our hope that some of the ideas presented in this paper will be
useful for these systems.

\acknowledgments

We devote this paper to the memory of Prof. I. Prigogine, who guided
us  in the initial stage of this work. We thank Dr. T. Petrosky and
Dr. E. Karpov for many suggestions and discussions. We acknowledge
the International Solvay Institutes for Physics and Chemistry, the
Engineering Research Program of the Office of Basic Energy Sciences
at  the U.S. Department of Energy, Grant No DE-FG03-94ER14465,  and
the Robert A. Welch Foundation Grant F-0365 for partial support of
this work.

\appendix

 \section{Coefficients of  transformed operators}
 \label{Coefficients}
 For the integrable case we have
 \ba \label{cd}
        && \cb_{11}=-\bar{N}\frac{\wb_1+\w_1}{2\w_1} \,,   \\
         &&     \cb_{1k}=\bar{N}\frac{\lam V_k}{\w_k-\wb_1}    \,,   \\
        &&  \db_{11}=-\bar{N}\frac{\wb_1-\w_1}{2\w_1} \,,   \\
         &&   \db_{1k}=-\bar{N}\frac{\lam V_k}{\w_k+\wb_1}\,,    \\
&& \cb_{kk} =  1 \,,   \\
&&  \cb_{k1} =  -\lambda V_k G^+(\omega_k)  (\ome_k + \ome_1) \,,   \\
&&  \db_{k1} =  -\lambda V_k G^+(\omega_k)  (\ome_k - \ome_1) \,,   \\
&&  \cb_{kk'} =  2\omega_1 \lambda V_k G^+(\omega_k) \frac {\lam V_{k'}}{\w_{k'}-\w_k- i\epsilon} \quad , \quad (k\ne k')\,,   \nonumber\\
\\
 && \db_{kk'} = -2\omega_1 \lambda V_k G^+(\omega_k) \frac {\lam V_{k'}}{\w_{k'}+\w_k}    \,.
 \ea
The normalization constant $\bar{N}$  given by 
\ba \label{Nb}
        \bar{N}^2=\frac{\w_1}{\wb_1}\left[1+\int_{0}^\infty dk\,
                \frac{4\w_1 \w_k \lam^2 v_k^2}{(\w_k^2-\wb_1^2)^2}\right]^{-1} \,.
\ea 

For the non-integrable case we have \cite{Antoniou01} 
\ba     \label{cdcom}
            c_{11}&=& -N^*\frac{z_1^*+\w_1}{2\w_1}       \,, \\
             c_{1k}&=& N^*\frac{\lam V_k}{(\w_k-z_1^*)_-}    \,,   \\
             d_{11}&=& -N^*\frac{z_1^*-\w_1}{2\w_1}    \,,  \\
            d_{1k}&=& -N^*\frac{\lam V_k}{\w_k+z_1^*}    \,, \\
 c_{k1} &=&  -\lambda V_k G_d^+(\omega_k)  (\ome_k + \ome_1) \,,   \\
 d_{k1} &=&   -\lambda V_k G_d^+(\omega_k)  (\ome_k - \ome_1) \,,   \\
c_{kk'} &=&   2\omega_1 \lambda V_k G_d^+(\omega_k) \frac {\lam V_{k'}}{\w_{k'}-\w_k- i\epsilon} \,, \quad (k\ne k')\,,   \no\\
 d_{kk'} &=&  -2\omega_1 \lambda V_k G_d^+(\omega_k) \frac {\lam V_{k'}}{\w_{k'}+\w_k}    \,.
 \ea
 and
\ba
 \ct_{k1} &=&   \cb_{k1},  \qquad   \ct_{kk'} =   \cb_{kk'} \,,   \\
 \dt_{k1} &=& \db_{k1},   \qquad    \dt_{kk'} =  \db_{kk'}   \,.
 \ea
We define $G_d^+(\w_k)$ as
\ba \label{Gt}
        G_d^+(\w_k)&\equiv& G^+(\w_k)-i \, \frac{\pi N^2}{ \w_1}
            \delta_c(\w_k-z_1)  \,,
\ea
where  $-N^2/(2 \w_1)$ is the residue of $G^+(\w)$ at the pole $z_1$
in the second sheet. The  normalization constant $N$ is
\ba     \label{N}
            N^2=\frac{\w_1}{z_1}\left[1+\int_{0}^\infty dk\, \frac{4 \w_1\w_k
                    \lambda^2 v_k^2}{(\w_k^2-z_1^2)_+^2}\right]^{-1}.
\ea

These coefficients give the ``out'' eigenoperators of the
Hamiltonian, $\At_k^\dg$, $\At_1^\dg$ and their Hermitian
conjugates. In previous publications (e.g. \cite{PPT}) the ``in''
states were used  to obtain decaying states for $t>0$ in the
Schr\"odinger picture. In this paper we consider observables in the
Heisenberg picture,  where the ``out'' operators are the ones that
decay for $t>0$.

\section{Commutation relation}
\label{app:A1A1}

In this Appendix we prove the commutation relation 
\begin{eqnarray}
          [\At_1, \At_1^\dg]=0  \,,
\label{DD2}
\end{eqnarray}
Using the explicit forms of $\At_1, \At_1^\dg$ we obtain 
\begin{eqnarray}
[\At_1, \At_1^\dagger]   = |N|^2 \left(\sum^{\infty}_{k=0}
|c_{1k}|^2 - \sum^{\infty}_{k=0}
|d_{1k}|^2+\frac{\tilde{\w_1}}{\w_1} \right)  \,.
\end{eqnarray}
We will show that the expression inside parenthesis vanishes, 
\ba \label{cd=0} \sum_{k}|c_{1k}|^2-\sum_{k}|d_{1k}|^2
+\frac{\tilde{w}_1}{w_1}=0 \,, \ea 

We know that $z_1=\wt_1-i\gam$ is the pole of $G^+(\w)$, defined in
(\ref{Gz}). Therefore 
\begin{eqnarray}
\label{z1}
     \w_1^2-z_1^2&=&\sum_{k=0}^{\infty}
    \frac{4w_1\w_k \lambda^2V^2_k}{(\w_k^2-z_1^2)_+}  \no
        &=&\sum_{k=0}^{\infty}
    \frac{2w_1 \lambda^2V^2_k}{(\w_k-z_1)_+}
     +\sum_{k=0}^{\infty}
    \frac{2w_1 \lambda^2V^2_k}{\w_k+z_1} \,.
\end{eqnarray}
Subtracting the complex conjugate expression  and dividing the
result
 by $2 w_1(z_1-z_1^*)$ we have
\begin{eqnarray}
    && \sum^{\infty}_{k=0} \frac{\lambda^2
V_k^2}{|\omega_k-z_1|^2_+} -\sum^{\infty}_{k=0} \frac{\lambda^2
V_k^2}{|\omega_k+z_1|^2} \no && =-\frac{z_1+z_1^*}{2 w_1}
=-\frac{\wt_1}{\w_1}\,,
\end{eqnarray}
which is equivalent to \Ep{cd=0}. This proves the desired expression
(\ref{DD2}).

>From Eqs. (\ref{noHil}) and (\ref{anoint}) we find that $[a_1,
\Bt_1^\dg] = [\sum_k D_k, \Bt_1^\dg]$. We then deduce the following
other commutation relations
\ba \label{crel2}
        && \left[\At_k,A_1^\dg\right] = 2\pi i \lam V_k (\w_k-z_1) \del_c(\w_k-z_1) \,, \no
        &&  \left[A_k,\At_{1}^\dg\right] = - 2\pi i \lam V_k \frac{G^+(\w_k)}{G^-(\w_k)} (\w_k-z_1^*) \del_c(\w_k-z_1^*). \no
 \ea
If the test functions contain singularities at $\w_k=z_1$ or
$\w_k=z_1^*$, then these commutators are non-vanishing.

\section{Proof of Analyticity}
\label{Analyticity}
In this section we verify  condition (C) on the $\Pito$ projector
for the non-integrable case. This means that 
 \ba
 \label{An0}
  \lim_{\lam\to 0}  {\Pito}^\dg(a_1^{\dg m+1} a_1^n) &=&\Po (a_1^{\dg m+1} a_1^n) = a_1^{\dg m+1} a_1^n   \,, \\
 \label{An02}
  \lim_{\lam\to 0}   {\Pito}^\dg(a^{\dg m}_1 a_1^{n+1})&=& \Po (a_1^{\dg m} a_1^{n+1}) = a_1^{\dg m} a_1^{n+1}  \,,
\ea 
for all $m,n\ge0$. Furthermore, $\Pito$ has to  be expandable in a
power series of $\lam$. If $\Pito$ satisfies these conditions we
will say, in short,  that it is analytic at $\lam=0$.  This property
is not trivial, because in \Ep{Pitomn} there appear non-analytic
terms in the products or  commutators of renormalized operators, as
in \Ep{noHil} (see also \cite{PPT}).

We have, using Eqs. (\ref{AntAk}) and (\ref{anoint}) 
\ba
  {\Pito}^\dg a_1^\dg = \Bt^\dg_1, \quad {\Pito}^\dg a_1 = \Bt_1.
\ea 
Both expressions are analytic at $\lam=0$. Assuming that
${\Pito}^\dg(a_1^{\dg m} a^n_1)$, ${\Pito}^\dg(a_1^{\dg m-1} a^n_1)$
and ${\Pito}^\dg(a_1^{\dg m} a^{n-1}_1)$ are analytic at $\lam=0$,
we will show  that the recursive expression (\ref{Pim+1n}) is also
analytic. This will prove \Ep{An0}  by recursion.

We start with the first term in the right hand side of Eq.
(\ref{Pim+1n}),
\ba \label{mn}
      &&  \Pito^\dg a_1^\dg \cdot {\Pito}^\dg(a_1^{\dg m} a^n_1) \\
      &=& \Bt^\dg_1  \cdot {\Pito}^\dg \left[ (\Bt^\dg_1+ \sum_k D_k^\dg)^m ( \Bt_1+ \sum_k D_k)^n
        \right]  \,. \nonumber
\ea
As we show now, this product generates non-analytic terms, even if
its two factors are analytic. In the products between $\Bt_1^\dg$
outside the brackets and  either $\Bt_1$  or $\Bt_1^\dg$  inside the
brackets there appears the term $\At_1^\dg \At_1$, which has the
following $\lam$ dependence 
\ba
  \At_1^\dg \At_1 &=& g_0 (\lam) a_1^\dg a_1 + \lam \int dk\,  g_1 (k, \lam) a_1^\dg a_k + \cdots \no
&+& \lam^2 \int dk\,  g_2(k, \lam) a_k^\dg a_k + \cdots     \,. \ea
For the perturbation expansion to exist, the functions $g_0(\lam)$,
$ g_1(k,\lam),   g_2(k,\lam), \cdots$ must be analytic at  $\lam=
0$, with $g_0(0)=1$. However, $ g_2(k, \lam)$ is not analytic at
$\lam=0$. We have
\ba \label{A1A1}
  &&  \lam^2 g_2(k,\lam)   \\
&=&  |N|^2   \lam^2 v_k^2 \frac{1}{(\w_k-z_1)_+}
\frac{1}{(\w_k-z_1^*)_-}  \no
 &=&  |N|^2  \frac{\lam^2 v_k^2}{z_1-z_1^*} \left[\frac{1}{(\w_k-z_1)_+} - \frac{1}{(\w_k-z_1^*)_-}  \right]   \,.\nonumber
\ea
For $\lam\to 0$  the term inside brackets goes to 
\ba
 \frac{1}{\w_k-\w_1 - i\eps} - \frac{1}{\w_k-\w_1 + i\eps}  = 2\pi i \del(\w_k - \w_1)  \,.
\ea 
Moreover we have (see Eq.  (\ref{gam})) 
\ba
 z_1-z_1^* =  -2 i \gamma =  - 2 \pi i \lam^2 \vt^2(\w_1)+ O(\lam^3)
\ea 
with 
\ba
 \label{vtdef}
 \vt^2(\w_k) \equiv v_k^2 \frac{dk}{d\w_k},.
\ea 
This leads to
\ba \label{A1A2}
  \lim_{\lam\to 0} \lam ^2 g_2(k,\lam)
 = -  \frac{ d\w_k}{dk} \del (\w_k-\w_1)
\ea
which is non-zero.

Coming back to \Ep{mn}, the term $\Bt_1^\dg$ outside brackets  in
the right hand side can pair with either   $m$ of the $\Bt_1^\dg$ or
$n$ of the $\Bt_1$  inside brackets. Thus all the non-analytic terms
involving $\At_1^\dg \At_1$ are included in
\ba
 \label{PiBPi}
        \left[
            {\Pito}^\dg a^\dg_1 \cdot {\Pito}^\dg( a_1^{\dg m} a^n_1)
                \right]_{non}
                &=&  n(\Bt^{\dg}_1\Bt_1)_{\rm non} \cdot {\Pito}^\dg (a_1^{\dg m} a^{n-1}_1) \no
                 &+& m ( \Bt^{\dg 2}_1 )_{\rm non} \cdot {\Pito}^\dg(a_1^{\dg m-1} a^n_1) \,. \no
\ea
In order for Eq. (\ref{Pim+1n}) to be analytic, the second and third
terms in the right hand side of Eq. (\ref{Pim+1n}) should cancel the
non-analytic terms of \Ep{PiBPi}.  Combining \Ep{Pim+1n} and
\Ep{PiBPi} we obtain
\ba
 \label{C8}
        \left[{\Pito}^\dg(a_1^{\dg m+1} a_1^n)\right]_{\rm non}  &=&
      n
    (\Bt^\dg_1 \Bt_1+\Yt )_{\rm non} \cdot {\Pito}^\dg(a^{\dg m}_1   a^{n-1}_1) \no
     &+& m(\Bt^{\dg 2}_1 + \Xt)_{\rm non} \cdot {\Pito}^\dg(a^{\dg m-1}_1 a^n_1) \,. \no
\ea

>From \Ep{A1A2} we get 
 \ba
  \lim_{\lam\to 0} (\Bt^\dg_1 \Bt_1)_{\rm non} =  -  \int_0^\infty d\w_k\, \del (\w_k-\w_1)a^\dg_k a_k
 \ea
where we used $\lim_{\lam\to 0} |c_{11}|^2 = 1$ and $\lim_{\lam\to
0} |d_{11}|^2 = 0$. On the other hand we have 
\ba \label{G+G+}
   Y_{\rm non} &=& \sum_k |\ct_{k1}|^2 a^\dg_k a_k  \\
  &=& \int_0^\infty dk\, \lam^2 v_k^2 |G^+(\w_k)|^2 (\w_k + \w_1)^2 a^\dg_k a_k \nonumber
\ea
where we replace $\At_k^\dg \At_k$ by $a^\dg_k a_k$ in the
thermodynamic limit (see \Ep{Akak}).

In the limit $\lam\to 0$ we obtain (see \Ep{G2''})
\ba \label{G+G+2}
   \lim_{\lam\to 0} Y_{\rm non} &=& \int_0^\infty d\w_k\, \frac{1}{4\w_1^2} \del(\w_k-\w_1) (\w_k + \w_1)^2 a^\dg_k a_k \no
           &=& \int_0^\infty d\w_k\,  \del(\w_k-\w_1) a^\dg_k a_k  \,.
\ea
Thus we get
\ba \label{BY}
   \lim_{\lam\to 0} (\Bt^\dg_1 \Bt_1+\Yt )_{\rm non}  = 0
\ea
which shows that the non-analytic terms cancel.

Similarly one can show that
\ba \label{BX}
   \lim_{\lam\to 0} (\Bt^{\dg 2}_1 +\Xt )_{\rm non}  = 0  \,.
\ea

Since we assume that ${\Pito}^\dg(a_1^{\dg m-1} a^n_1)$ and
${\Pito}^\dg(a_1^{\dg m} a^{n-1}_1)$ are analytic,  we conclude that
(\ref{Pim+1n}) is analytic at $\lam=0$. Thus, by recursion $
{\Pito}^\dg(a_1^{\dg m+1} a_1^n)$ is analytic for arbitrary $m,n$.

We can show in the same way that the recursive expression
(\ref{Pitmn-})  is analytic at  $\lam=0$, which proves \Ep{An02}.

\section{Evaluating $\wt_1$ and $\lam$ to $O(\lam^2)$}
\label{app:z1}

>From \Ep{z1} we have
\ba   \w^2_1-z^2_1-\int_{0}^\infty dk\, \frac{4 \w_1 \w_{k} \lam^2
v^2_{k}}{(\w^2_{k}-z^2_1)_+}=0   \,. \ea
Approximating $z^2_1\simeq\w^2_1 +i\eps$ in the denominator we have
\ba \label{B2}
        \w^2_1-\wt^2_1+2i\wt_1 \gam - \int_{0}^\infty dk\,
    \frac{4 \w_1 \w_{k} \lam^2 v^2_{k}}{\w^2_{k}-\w^2_1-i\eps}=0
    \,.
\ea

Writing
\ba &&  \frac{1}{\w^2_k-\w^2_1 -i\eps}= {\cal
P}\frac{1}{\w^2_k-\w^2_1}
       +i\pi \delta(\w^2_k-\w^2_1)        \\
       &=&{\cal P}\frac{1}{\w^2_k-\w^2_1} +\frac{i\pi}{2\w_1} \delta(\w_k-\w_1)   \,,\nonumber
\ea
we then obtain for the real part and imaginary part of (\ref{B2}) as
\ba \label{wt}
      \wt^2_1=\w^2_1-{\cal P}\int_{0}^\infty dk\, \frac{4 \w_1\w_k \lam^2
      v^2_k}{\w^2_k-\w^2_1}+O(\lam^4)     \,.
\ea
Therefore,
\ba    \label{wapprox}
    \wt_1 &\simeq& \w_1 - {\cal P}\int_{0}^\infty dk\, \frac{ 2 \w_k \lam^2
    v_k^2 }{ \w_k^2 - \w_1^2 }  \,.
\ea
and (see \Ep{vtdef})
\ba \label{gam}
   \gam=\pi \lam^2 \vt^2(\w_1) + O(\lam^4)   \,.
\ea

\section{Green's function in the weak-coupling approximation}
\label{pole}

In this Appendix we  find an approximation  \ of Green's function
$G^+$ valid for weak coupling. We start by expanding the inverse
function around the pole $z_1$.  Since this function depends on
$\w_k^2$, i.e.,
\ba \label{Gk}
    [G^+(\ome_k^2)]^{-1}=\w^2_1-\w^2_k-\int_{0}^\infty dk'\,
        \frac{4 \w_1 \w_{k'} \lam^2
        v^2_{k'}}{\w^2_{k'}-\w^2_k-i\eps}\,,
\ea
we make an expansion in the variable $\w_k^2$ around $z_1^2$,
\ba && G^+(\ome_k^2)^{-1} = G^+(z_1^2)^{-1} + (\ome_k^2-z_1^2)
[G^+(z_1^2)^{-1}]' \no &+&  \frac{1}{2} (\ome_k^2-z_1^2)^2
[G^+(z_1^2)^{-1}]'' + \cdots   \,. \ea
We have  $G^+_k (z_1^2)^{-1} =0$. The first derivative term is given
by  (with $N$ defined in Eq. (\ref{N}))
\ba \frac{d[G^+(\ome_k^2)]^{-1}}{d\w^2_k}\bigg|_{\w^2_k=z^2_1} &=&
        -\left(1+\int_{0}^\infty dk'\, \frac{4 \w_1\w_{k'} \lam^2
        v^2_{k'}}{(\w^2_{k'}-z^2_1)^2}\right)  \no
        &=& -\frac{\w_1}{z_1} \frac{1}{N^2}    \,.
\ea        
For weak coupling we may neglect the second and higher derivative
terms. This gives
\ba     G^+(\ome_k)= -\frac{z_1}{\w_1}
        \frac{N^2}{\w^2_k-z^2_1} + {\rm higher \, derivatives}  \,.
\ea
Furthermore for weak coupling we have $N = 1 + O(\lam^2)$, $z_1 =
\w_1 + O(\lam^2)$. Thus we get
\ba \label{G+k}
      G^+(\ome_k) &\simeq& \frac{1}{(z_1-\w_k)(z_1+\w_k)}  \no
            &\simeq& \frac{1}{(\w_1-\w_k -i\eps)(\w_1+\w_k)}   \no
            &=& \frac{1}{\w_1^2-\w_k^2 -i\eps}   \no
        &=& -{\cal P}\frac{1}{\w^2_k-\w^2_1} +\frac{i\pi}{2\w_1}\delta(\w_k-\w_1)
\ea
and similarly,
\ba  \label{G-k}
        G^-(\ome_k)
        &\simeq& \frac{1}{\w_1^2-\w_k^2 +i\eps}   \no
        &=&     -{\cal P}\frac{1}{\w^2_k-\w^2_1} - \frac{i\pi}{2\w_1}\delta(\w_k-\w_1) \,.
\ea
Another useful formula follows from the exact relation
\ba \label{G2'}
       4 \pi i \lam^2 \vt^2(\w_k)   \w_1 |G^+(\ome_k)|^2    = G^+(\w_k) -  G^-(\w_k)
\ea
where $\vt^2(\w_k) \equiv v_k^2\, dk/d\w_k.$ Using (\ref{G+k}) and
(\ref{G-k}), we find that
\ba \label{G-G}
        G^+(\ome_k) - G^-(\ome_k) \simeq \frac{i\pi}{\w_1} \delta (\w_k-\w_1)
         \,.
\ea
Combining this result with \Ep{G2'} we get
\ba \label{G2''}
       4  \lam^2 \vt^2(\w_k) |G^+(\ome_k)|^2    \simeq  \frac{1}{\w_1^2} \delta (\w_k-\w_1)
\ea
in the lowest order approximation in $\lam$ expansion.

\section{Trace relations involving  $a^\dg_1$ and
$a_1$} \label{AppC}

Using the following relationships,
\ba     \left[
{a_1},{a^{\dg n}_1}\right]&=& n a^{\dg n-1}_1    \,,   \\
         \left[   {a}^\dg_1,{a^m_1} \right]&=&   -m a^{m-1}_1  \,,   \\
              \left[a^\dg_1 a_1,a^{\dg n}_1\right] &=&  n a^{\dg n}_1  \,,   \\
                  \left[a^\dg_1 a_1,a^{ m}_1\right] &=& -m a^{m}_1  \,,   \\
             \left[a^{\dg 2}_1,a^{\dg n}_1 a^{ m}_1\right] &=& -2m a^{\dg n+1}_1 a^{ m-1}_1
                -m(m-1) a^{\dg n}_1 a^{ m-2}_1   \,,   \no\\
                 \left[a^{ 2}_1,a^{\dg n}_1 a^{ m}_1\right] &=& 2n a^{\dg n-1}_1 a^{ m+1}_1
                +n(n-1) a^{\dg n-2}_1 a^{ m}_1   \,,\no
\ea
we can show that
\begin{widetext}
\begin{align} \label{eq}
   (n-m) {\rm Tr} \left[ a^{\dg n}_1 a^{ m}_1 \rhoo\right]
            &= {\rm Tr} \left[ [ a^\dg_1 a_1,a^{\dg n}_1 a^{ m}_1] \rhoo \right]  \,,
                           \\
   (n+m) {\rm Tr} \left[ a^{\dg n}_1 a^{ m}_1
                \rhoo\right]
            &= {\rm Tr} \left[ \{a^\dg_1 a_1,a^{\dg n}_1 a^{ m}_1\} \rhoo \right]
              -2 {\rm Tr} \left[ a^{\dg n}_1 a^{ m}_1 a_1 \rhoo a^\dg_1\right]  \,,
                        \\
   n m {\rm Tr} \left[ a^{\dg n-1}_1 a^{m-1}_1 \rhoo\right]
             &= {\rm Tr} \left[ a^{\dg n}_1 a^{ m}_1 a^\dg_1 \rhoo a_1\right]
                 -(n+m+1) {\rm Tr} \left[  a^{\dg n}_1 a^{m}_1  \rhoo\right]
                 -    {\rm Tr} \left[ a^{\dg n}_1 a^{ m}_1 a_1 \rhoo a^\dg_1\right]  \,,   \\
   {\rm Tr} \left[ m a^{\dg n+1}_1 a^{m-1}_1 \rhoo \right]
             &= -{\rm Tr} \left[a^{\dg n+1}_1 [a^\dg_1,a^m_1]\rhoo \right]
                         = {\rm Tr} \left[ a^{\dg n}_1 a^{ m}_1 (a^\dg_1\rhoo a^\dg_1-\rhoo a^\dg_1 a^\dg_1)\right] \,,   \\
 {\rm Tr} \left[ n a^{\dg n-1}_1 a^{m+1}_1 \rhoo\right]
          &= {\rm Tr} \left[  [a^\dg_1,a^n_1]a^{\dg m+1}_1\rhoo\right]
                       = {\rm Tr} \left[ a^{\dg n}_1 a^{ m}_1 (a_1\rhoo a_1-a_1 a_1\rhoo)\right] \,, \\
   {\rm Tr} \left[ m(m-1)a^{\dg n}_1 a^{m-2}_1 \rhoo\right]
          &= -{\rm Tr} \left[2m a^{\dg n+1}_1 a^{m-1}_1 \rhoo\right]
            -{\rm Tr} \left[  [a^\dg_1 a^\dg_1 ,a^{\dg n}_1 a^{m}_1 ] \rhoo\right]
              = -{\rm Tr} \left[ a^{\dg n}_1 a^{m}_1 (2a^\dg_1\rhoo
            a^\dg_1-a^\dg_1 a^\dg_1\rhoo-\rhoo a^\dg_1
            a^\dg_1)\right]  \,, \\
   {\rm Tr} \left[ n(n-1)a^{\dg n-2}_1 a^{m}_1 \rhoo\right]
             &= -{\rm Tr} \left[2n a^{\dg n-1}_1 a^{m+1}_1  \rhoo \right]
            +{\rm Tr} \left[  [a_1 a_1 ,a^{\dg n}_1 a^{m}_1 ] \rhoo\right]
                                 = -{\rm Tr} \left[ a^{\dg n}_1 a^{m}_1 (2a_1\rhoo
            a_1-a_1 a_1\rhoo-\rhoo a_1
            a_1)\right]  \,.
\end{align}
Furthermore,
\ba
        && a^\dg_1 a_1 \rhoo  + \rhoo
                a^\dg_1 a_1
            -a_1\rhoo  a^\dg_1-a^\dg_1 \rhoo a_1
            +\rhoo   = \frac{1}{2} (a^\dg_1 a_1 \rhoo + \rhoo a^\dg_1 a_1
            -2 a_1\rhoo a^\dg_1+a_1 a^\dg_1 \rhoo +
            \rhoo a_1 a^\dg_1-2 a^\dg_1 \rhoo a_1)
            \,.
\ea
\end{widetext}

\section{Coordinate representation of $a^\dg_1$ and $a_1$}
\label{AppD}

Starting from
\ba
        a^\dg_1
         =\sqrt{\frac{M_1 \w_1}{2}} ({q_1}-i\frac{{p_1}}{M_1 \w_1})
          \quad {\rm and} \quad
          {p_1}=\frac{1}{i} \frac{\pdi}{\pdi {q_1} } \,,
\ea
for an arbitrary vector $\phr$, we find that
\ba \label{adx}
          \ql a^\dg_1 \phr&=& \frac{1}{\sqrt{2}} \left(
          x_1-\frac{\pdi}{\pdi x_1}\right) \ql\rph   \,, \\
             \ql a_1 \phr&=& \frac{1}{\sqrt{2}} \left(
          x_1+\frac{\pdi}{\pdi x_1}\right) \ql\rph   \,, \\
             \phl a^\dg_1 \qpr&=& \frac{1}{\sqrt{2}} \left(
          {x'}_1+\frac{\pdi}{\pdi {x'}_1}\right) \lph\qpr   \,, \\
          \phl a_1 \qpr&=& \frac{1}{\sqrt{2}} \left(
          {x'}_1-\frac{\pdi}{\pdi {x'}_1}\right) \lph\qpr   \,.
\ea
The ket of the dimensionless coordinate $x_1$ is related to $q_1$ by
$|q_1\>=\left(M_1 \w_1 \right)^{1 / 4 }|x_1\>$. From the relation
$a_1 a^\dg_1-a^\dg_1 a_1=1 $, we also find that
\ba         \ql a_1a^\dg_1 \phr&=& \frac{1}{2}\left(
          x_1+\frac{\pdi}{\pdi x_1}\right) \left(
          x_1-\frac{\pdi}{\pdi x_1}\right) \ql\rph,  \no
             \ql a^\dg_1a_1 \phr&=& \frac{1}{2} \left(
          x_1-\frac{\pdi}{\pdi x_1}\right)\left(
          x_1+\frac{\pdi}{\pdi x_1}\right) \ql\rph, \no
     \phl a_1 a^\dg_1  \qpr
       &=&\frac{1}{2}\left({x'}_1+\frac{\pdi}{\pdi {x'}_1}\right)
          \left({x'}_1-\frac{\pdi}{\pdi {x'}_1}\right) \lph\qpr, \no
     \phl  a^\dg_1 a_1 \qpr&=&\frac{1}{2}\left({x'}_1-\frac{\pdi}{\pdi {x'}_1}\right)\left(
          {x'}_1+\frac{\pdi}{\pdi {x'}_1}\right) \lph\qpr. \no
\ea
We then deduce that
\begin{widetext}
\begin{align}
        & \ql[a^\dg_1 a_1, \rhoo]\qpr
               = \frac{1}{2}\left[ -
                            \left(\frac{\pdi^2}{\pdi x^2_1}
                     -\frac{\pdi^2}{\pdi {x'}^2_1}\right) +(x^2_1-{x'}^2_1)\right]\rhoo(x_1,x'_1)    \,,  \\
    &  \ql ([a_1\rhoo,a^\dg_1]+[a_1,\rhoo a^\dg_1]
                +[a^\dg_1\rhoo ,a_1]+[a^\dg_1,\rhoo a_1])\qpr =  \left[ -(x_1-{x'}_1)^2 + \left( \frac{\pdi}{\pdi {x}_1}+\frac{\pdi}{\pdi {x'}_1}\right)^2\right]
                        \rhoo (x_1,{x'}_1)    \,,   \\
   &  \ql ([ a_1\rhoo ,a^\dg_1]+[a_1,\rhoo a^\dg_1]
             -[a^\dg_1\rhoo ,a_1]-[a^\dg_1,\rhoo a_1]) \qpr  \no
       & \qquad\qquad\qquad =\bigg[ -(x_1-{x'}_1)\left( \frac{\pdi}{\pdi {x}_1}-\frac{\pdi}{\pdi {x'}_1}\right)
          + \left( \frac{\pdi}{\pdi {x}_1}+\frac{\pdi}{\pdi {x'}_1}\right)(x_1+{x'}_1)\bigg]
                 \rhoo (x_1,x'_1)   \,,   \\
    &  \ql  ( [a^\dg_1\rhoo ,a^\dg_1]+[a^\dg_1,\rhoo a^\dg_1]
              +[a_1\rhoo ,a_1]+[a_1,\rhoo a_1] )\qpr
         =  -\left[(x_1-{x'}_1)^2 + \left( \frac{\pdi}{\pdi {x}_1}
             +\frac{\pdi}{\pdi {x'}_1}\right)^2\right]\rhoo (x_1,x'_1)  \,,   \\
      & \ql (  [a^\dg_1\rhoo ,a^\dg_1]+[a^\dg_1,\rhoo a^\dg_1]
             -[a_1\rhoo ,a_1]-[a_1,\rhoo a_1] ) \qpr
          = 2(x_1-{x'}_1)\left( \frac{\pdi}{\pdi {x}_1}+\frac{\pdi}{\pdi {x'}_1}\right)\rhoo (x_1,x'_1)   \,,   \\
      & \ql (  [ a^\dg_1,\rhoo a^\dg_1]+[a_1\rhoo ,a_1] )\qpr \\
        & \qquad\qquad      =-\frac{1}{2}  \bigg[(x_1-{x'}_1)^2 + \left( \frac{\pdi}{\pdi {x}_1}
             +\frac{\pdi}{\pdi {x'}_1}\right)^2  +  (x_1-{x'}_1)\left( \frac{\pdi}{\pdi {x}_1}-\frac{\pdi}{\pdi {x'}_1}\right)
           + \left( \frac{\pdi}{\pdi {x}_1}+\frac{\pdi}{\pdi {x'}_1}\right)(x_1+{x'}_1 )  \bigg]
                 \rhoo (x_1,x'_1)   \,, \no
       &\ql (  [ a^\dg_1,\rhoo a^\dg_1]- [a_1\rhoo ,a_1] )\qpr
            =\frac{1}{2}\left[ (x^2_1-{x'}^2_1)+\left(\frac{\pdi^2}{\pdi x^2_1}
          -\frac{\pdi^2}{\pdi {x'}^2_1}\right)+2(x_1-{x'}_1)\left( \frac{\pdi}{\pdi {x}_1}
          +\frac{\pdi}{\pdi
          {x'}_1}\right)\right]\rhoo(x_1,x'_1)\,.
\end{align}
\end{widetext}

\end{document}